\PassOptionsToPackage{x11names,table,dvipsnames}{xcolor}
\documentclass[10pt,twocolumn,letterpaper]{article}

\usepackage[pagenumbers]{cvpr} 

\usepackage{bm}
\usepackage{makecell}
\usepackage{booktabs}

\def\paperID{7727} 
\def\confName{CVPR}
\def\confYear{2025}
\newcommand{\METHOD}{DecoupledGaussian}

\title{\METHOD: Object-Scene Decoupling for Physics-Based Interaction}

\author{Miaowei Wang$^{1}$\quad Yibo Zhang$^{2}$  \quad Rui Ma$^{2}$\quad Weiwei Xu$^{3}$ \quad Changqing Zou$^{3}$ \quad Daniel Morris$^{4}$ \\
$^{1}$ The University of Edinburgh, $^{2}$ Jilin University, $^{3}$ Zhejiang University, $^{4}$ Michigan State University
}

\begin{document}

\twocolumn[{%
\renewcommand\twocolumn[1][]{#1}%
\maketitle
\begin{center}
    \centering
    \captionsetup{type=figure}
    \includegraphics[width=\textwidth, trim=0 0 0 0, clip]{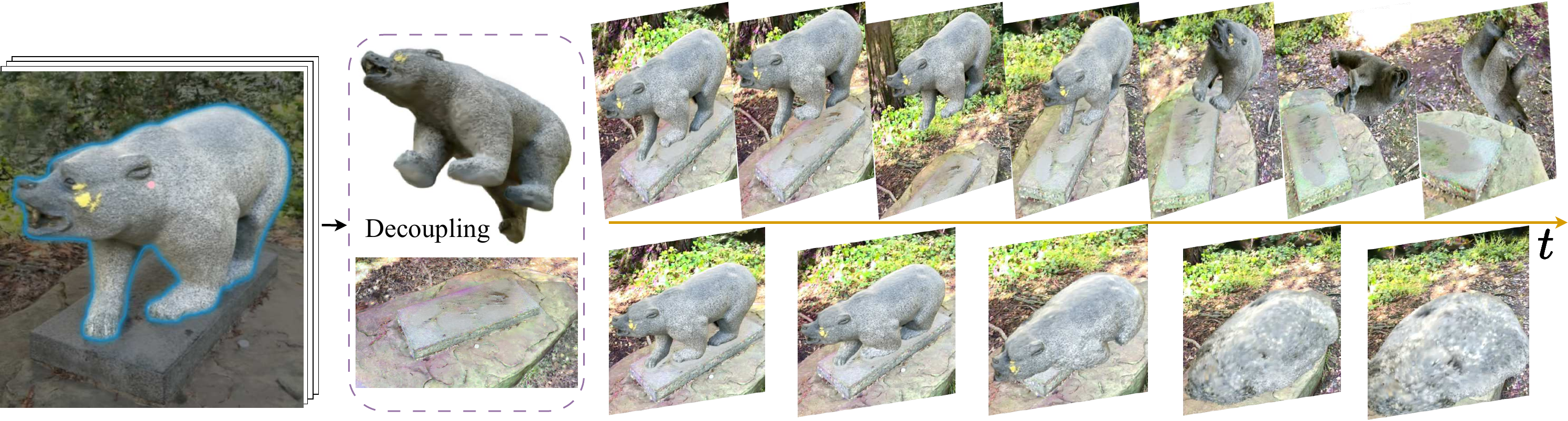}
    \captionof{figure}{\textbf{DecoupledGaussian} decomposes static objects and contacted scenes from videos or multi-view images, enabling simulations like scene collisions (Top) and object melting with material adjustments (Bottom). See the supplementary video for the full sequences.}
\label{fig:teaser}
\end{center}
}]

\begin{abstract}
We present DecoupledGaussian, a novel system that decouples static objects from their contacted surfaces captured in-the-wild videos, a key prerequisite for realistic Newtonian-based physical simulations. Unlike prior methods focused on synthetic data or elastic jittering along the contact surface, which prevent objects from fully detaching or moving independently, DecoupledGaussian allows for significant positional changes without being constrained by the initial contacted surface. Recognizing the limitations of current 2D inpainting tools for restoring 3D locations, our approach proposes joint Poisson fields to repair and expand the Gaussians of both objects and contacted scenes after separation. This is complemented by a multi-carve strategy to refine the object's geometry. Our system enables realistic simulations of decoupling motions, collisions, and fractures driven by user-specified impulses, supporting complex interactions within and across multiple scenes. We validate DecoupledGaussian through a comprehensive user study and quantitative benchmarks. This system enhances digital interaction with objects and scenes in real-world environments, benefiting industries such as VR, robotics, and autonomous driving. Our project page is at: \url{https://wangmiaowei.github.io/DecoupledGaussian.github.io/}.
\end{abstract}
\section{Introduction}
Interactive reconstruction and simulation of target objects and their surrounding scenes have become increasingly sophisticated recently. These can provide 4D assets for autonomous driving \cite{rong2020lgsvl} and robotics \cite{mittal2023orbit}, and also enable immersive applications in Virtual Reality (VR) \cite{moussa2022effectiveness} and the entertainment industry \cite{yeh2006enabling}.

Advances in realism have been made by moving beyond traditional representations, such as point clouds \cite{klein2004point}, meshes \cite{pfaff2021learning}, grids \cite{cohen2010interactive}, and signed distance fields \cite{seyb2019non}. Neural Radiance Fields (NeRF) \cite{mildenhall2021nerf} use neural rendering techniques for novel view synthesis from videos, enabling interactive games \cite{10.1145/3635705}, animation \cite{yin2023nerfinvertor}, and simulations \cite{li2023pacnerf}, where \emph{what is simulated directly stems from what is captured}. And Gaussian Splatting (GS) \cite{kerbl20233d}, known for its rapid rendering and reconstruction speeds, leverages discrete Gaussian kernels, making it easier to directly manipulate and process \cite{guedon2024sugar, chen2024gaussianeditor} objects reconstructed from videos.

However, current physics-based simulation methods that use NeRF \cite{li2023pacnerf,feng2024pie} or Gaussian splatting \cite{xie2024physgaussian,jiang2024vr,cai2024gaussian} either focus on synthetic objects, allowing for full-view observations during reconstruction, or simulate elastic deformations and jittering, in which objects remain constrained to the contacted surface. This prevents objects from truly detaching under user-specified impulses.

To allow objects to move without being constrained by the initial contacted surface, we need to  decouple objects from the contacted surface before simulation. In real-life settings, objects are influenced by gravity and typically rest on other surfaces, such as the sculpture on the pedestal in \cref{fig:teaser}. During imaging, an object and its contacted surfaces will be joined, resulting in hidden parts and occlusions and a fragmented representation of the object's surface. The primary challenge in decoupling, therefore, is accurately restoring and completing the 3D structures of both the object and its surrounding scene before simulation.

To tackle this challenge, we introduce \textbf{DecoupledGaussian}, a system that restores 3D geometry and textures of objects and contacted surfaces from in-the-wild videos using GS, laying the groundwork for realistic object-scene interactive simulations (see \cref{fig:teaser}). Notably, 2D inpainting (\cref{fig: inpainting_inconsistent}) often struggles with 3D restoration, especially in accurately capturing geometric positions. Our approach overcomes this by leveraging geometric priors assuming closed surfaces and multi-view observations from training viewpoints to restore realistic object and scene geometry.

Our method employs joint Poisson fields to reconstruct shape indicators for objects and scenes, resolving intersecting regions. Using Gaussian centers directly can introduce surface deviations due to blended rendering, causing artifacts in object reconstruction. To avoid this, we use unbiased depth maps from planar-based GS to create proxy points for realistic object reconstruction and reduce the scene's floaters through geometry regularization with flattened 3D Gaussians. To alleviate geometry expansion in Poisson reconstruction, we introduce a unilateral negative cross-entropy (UNCE) method for multi-view carving, refining the geometry to align with the observed views.

DecoupledGaussian is the first to restore both object and contacted surface geometry independently of 2D inpainting, which we use only for texture properties refinement. Extensive experiments on real-world videos, a new decoupling benchmark, user studies, quantitative comparisons, and ablations demonstrate our approach’s effectiveness in restoring accurate 3D properties and enabling precise interactive simulations. In summary, our
contributions include:
\begin{itemize} 
\item Development of an object-scene interactive simulation system that allows objects to detach from their contacted surfaces when constructed from in-the-wild videos and represented using GS. 
\item Introduction of geometric priors via joint Poisson fields and multi-view observations with UNCE for more realistic restoration (see \cref{fig: Qualitative_Comparision}) of geometric properties in GS.
\end{itemize}
\section{Related Work}
\paragraph{\emph{GS Editing}} A variety of methods have been proposed to modify or edit scenes built by Gaussian Splatting. \citet{wu2024deferredgs} enhance GS textures with learnable lighting adjustments, while \citet{fiebelman20244legs4dlanguageembedded} refine 4D video playback using human language prompts. Texture-GS \cite{xu2025texture} supports texture modifications via UV mapping decoupled from the original GS, and \citet{ma2024reconstructing} introduce deformations by aligning Gaussians to a proxy mesh with as-rigid-as-possible regularization. Additionally, GaussianEditor \cite{chen2024gaussianeditor} allows object addition and removal in GS scenes through 2D segmentation \cite{kirillov2023segment} and inpainting techniques \cite{suvorov2022resolution}. SC-GS \cite{huang2024sc} enables object deformation using sparse control points learned from dynamic video data, while \citet{modi2024simplicits} learns skinning weights for elastic deformations. \citet{huang2024gsdeformer} apply a bounding cage as a control proxy to deform GS representations.
\vspace{-4mm}
\paragraph{\emph{GS Simulation}} Gaussian Splatting can be incorporated into traditional simulation frameworks. For instance, PhysGaussian \cite{xie2024physgaussian} uses the Material Point Method (MPM) \cite{klar2016drucker,jiang2016material,stomakhin2013material} to simulate Gaussian kernels motion directly, while VR-GS \cite{jiang2024vr} applies eXtended Position-based Dynamics (XPBD) \cite{10.1145/2994258.2994272} to control GS via a bounding mesh. \citet{feng2024splashing} combine XPBD to model interactions between liquids and solids in GS, and \citet{borycki2024gasp} utilize MPM with triangle soups derived from GS. Additionally, \citet{abouphysically} apply GS in robotic decision-making through XPBD. For accurate physical property estimation in these simulations, GIC \cite{cai2024gaussian} derives physical properties from multiview video captures, building on techniques from \citet{li2023pacnerf,guan2022neurofluid} for system identification. Besides, \citet{liu2024physics3d} and \citet{huang2024dreamphysics} estimate material properties from synthetic video generated from static images using generative models. \citet{whitney2024learning} developed simulators trained on dynamic multi-view RGB-D video, and \citet{qiu-2024-featuresplatting} use visual language models to classify objects as elastic or rigid for text-driven physics simulations. However, these methods do not address the challenge of simulating an object detached from the contact surface when a user-provided impulse is applied.
\begin{figure}[!htb]
    \centering
    \vspace{-0.4cm}
    \includegraphics[width=\linewidth]{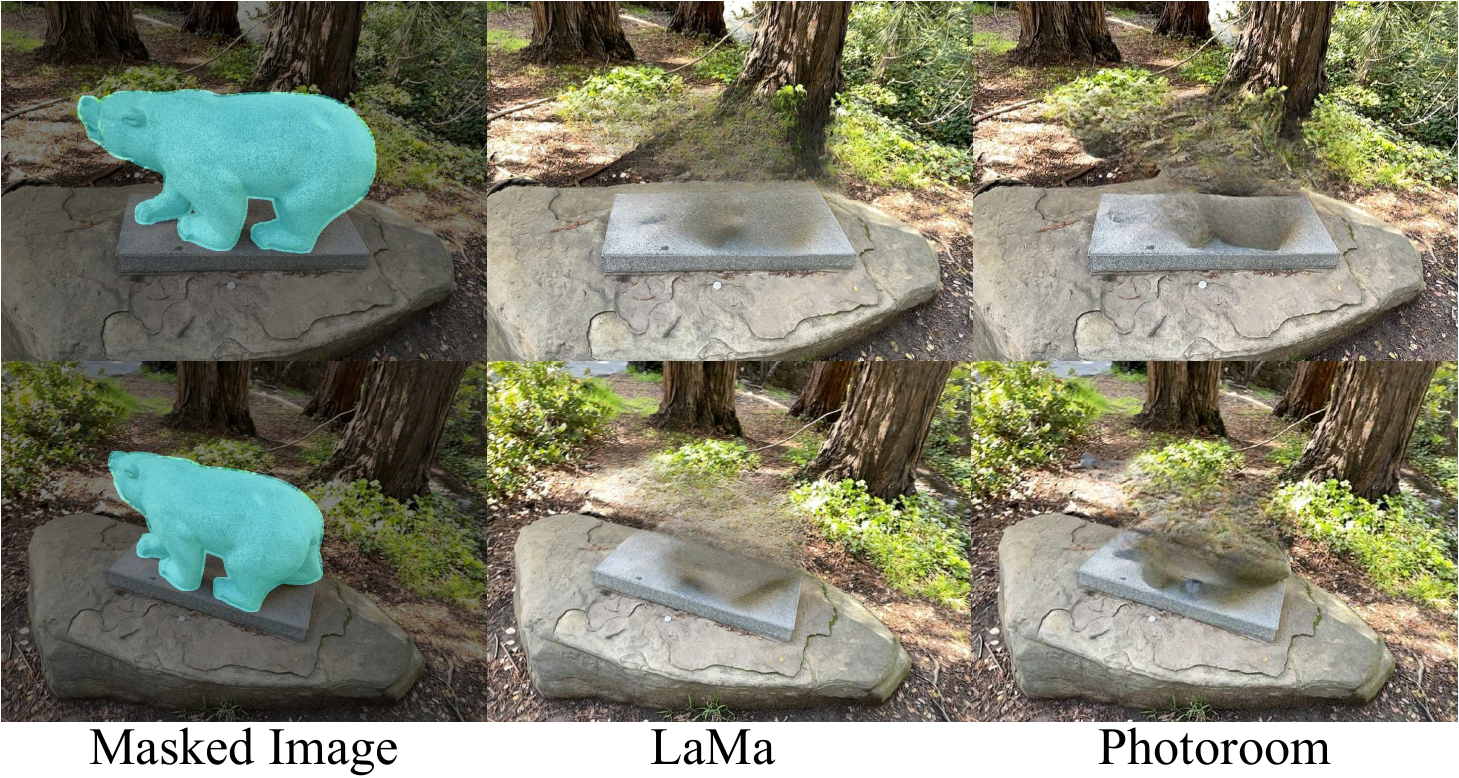}
    \vspace{-0.7cm}
    \caption{Inpainting tools (LaMa \cite{suvorov2022resolution}; PhotoRoom \cite{photoroom_remove_object}) introduce artifacts and inconsistent textures across frames.}
    \vspace{-0.4cm}
    \label{fig: inpainting_inconsistent}
\end{figure}
\vspace{-4mm}
\paragraph{\emph{GS Restoration}} Restoration techniques have addressed occluded mesh completion \cite{hu20242}, single-view depth point cloud completion \cite{wang2021unsupervised,huang2021rfnet,kasten2024point,huang2024zero,wu2024self}, and SDF-based reconstructions \cite{lu2018surface,gotsman2024linear,huangneuralindicator,peng2021shape}. In the context of Gaussian Splatting,  
 recent work has focused on restoring scene surfaces after object removal \cite{liu2024infusion, wang2024gscream} and reconstructing GS objects from sparse views \cite{yang2024gaussianobject}. However, a key challenge remains unaddressed: no GS or NeRF-based methods currently restore objects occluded by the scene or missed due to limited training viewpoints—a challenge mesh restoration techniques have begun to tackle \cite{hernandez20243d}. Similarly, video generation methods like PhysGen \cite{liu2025physgen}, which rely on static cameras, also overlook this issue. Current GS scene restoration techniques \cite{mirzaei2024reffusion, liu2024infusion, wang2024gscream} depend heavily on 2D inpainting tools \cite{suvorov2022resolution, podellsdxl, photoroom_remove_object} to fill gaps in geometry and texture post-object removal. However, these methods face two major issues (\cref{fig: inpainting_inconsistent}): (1) inpainted regions often fail to blend seamlessly with surrounding geometry, creating artifacts, and (2) texture inconsistencies across frames due to the lack of robust video inpainting tools. Our approach addresses these limitations by prioritizing geometry restoration, leveraging intrinsic GS geometry priors to ensure a coherent surface even when texture inpainting is imperfect.

\section{Preliminaries}
\subsection{3D Gaussian Splatting}
Gaussian Splatting \cite{kerbl20233d} represents a 3D scene with possible features by constructing 3D Gaussian kernels $\{\boldsymbol k_g, \sigma_g, \boldsymbol \Sigma_g, \mathcal{C}_g\}_{g \in \mathcal{G}}$, where $\boldsymbol k_g$, $\sigma_g$, $\boldsymbol \Sigma_g$, and $\mathcal{C}g$ denote Gaussian centers, opacities (encoding density), covariance matrices, and spherical harmonic (SH) representing color coefficients, respectively. The covariance matrix $\boldsymbol \Sigma_g$ at a Gaussian $g$ is factorized as $\boldsymbol \Sigma_g = \boldsymbol R_g \boldsymbol S_g\boldsymbol S_g^T \boldsymbol R_g^T$, where $\boldsymbol R_g$ is a rotation matrix, and corresponding scaling $\boldsymbol S_g= diag(s_1,s_2,s_3)$ is a diagonal matrix. Like NeRF \cite{mildenhall2021nerf}, GS is optimized for novel-view synthesis, where for a given 2D image plane, an integrated quantity $\boldsymbol q$ at a pixel $p$ is obtained by the following front-to-back $\alpha$-blending:
\vspace{-0.3cm}
\begin{equation}
\vspace{-0.3cm}
\boldsymbol q(p) = \sum_{i \in \mathcal{G}} \boldsymbol q_i \alpha_i \left[\prod_{j=1}^{i-1} (1-\alpha_j)\right]
\label{eq:GS-blending}
\end{equation}
where $\boldsymbol q_i$ is the quantity (for instance, SH-evaluated color $c_i$), and $\alpha_i$ is the termination probability derived from opacity $\sigma_i$ and affine-projected 2D Gaussian weights from $\boldsymbol \Sigma_i$.

\subsection{Continuum Simulation} 
We use the MLS-MPM \cite{hu2018moving} framework to solve Gaussian kernel governing equations (mass and momentum conservation) \cite{xie2024physgaussian}. The continuum is discretized into Lagrangian particles \( \mathrm{p} \), with time steps of \( \Delta t \) for deformation. At each step, particle mass and momentum are transferred to an Eulerian grid (P2G), where momentum is updated using the first Piola-Kirchhoff stress (PK1), and velocities \( v \) are advanced via forward Euler integration (Grid Operation). These grid velocities are then interpolated back to particles (G2P) for position updates during advection. MLS-MPM employs affine \( C_{\mathrm{p}} \) as a first-order approximation of \( \nabla v \), optimizing computation time. The elastic deformation gradient \( F^E \) is updated as \( F_\mathrm{p}^{n+1} = (I + \Delta t \, C_\mathrm{p}^n) F_\mathrm{p}^n \). Material parameters such as Young’s modulus \( E \) and shear modulus \( \mu \) \cite{jiang2016material} influence PK1 in grid momentum updates.

\begin{figure*}[!ht]
\centering
\includegraphics[width=\linewidth]{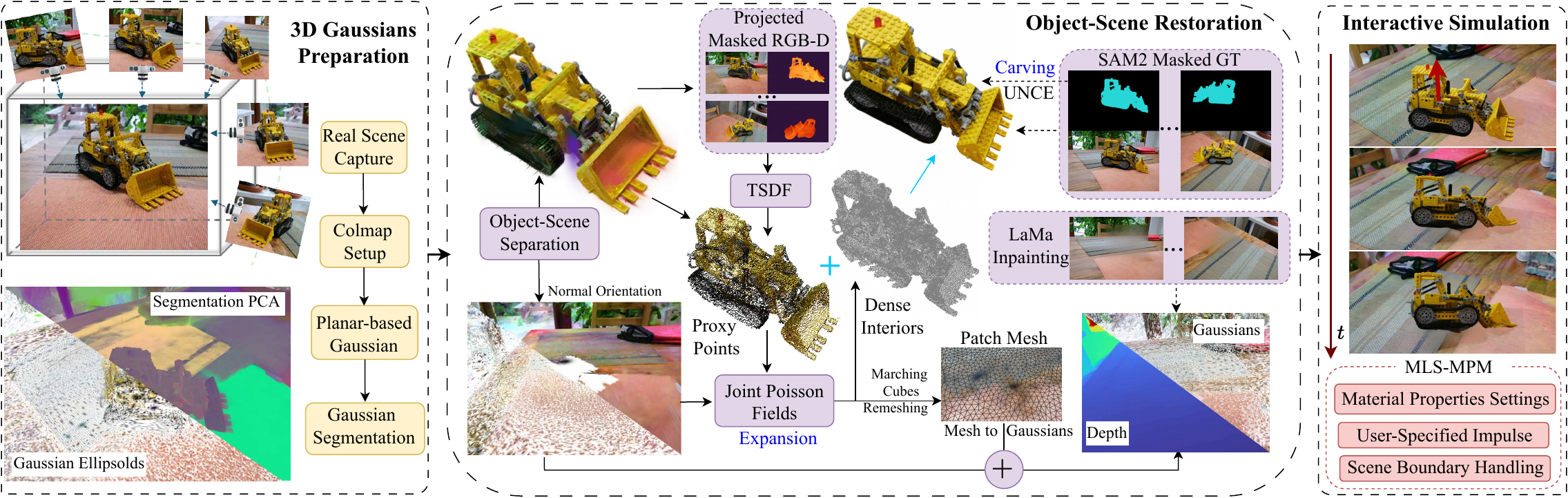}
\caption{\textbf{System Overview}. DecoupledGaussian is an interactive simulation system that enables objects to detach from their initial contact surfaces after applying our proposed restoration pipeline,   driven by user-specified impulses (red arrow on the right).}
\vspace{-4mm}
\label{fig: pipeline}
\end{figure*}

\section{DecoupledGaussian}
\label{sec:overview}
The DecoupledGaussian system starts with a reconstructed GS scene and allows an object resting on a planar surface to be moved off its surface in a realistic manner as
shown in \cref{fig: pipeline}.  First an object is segmented and a planar-based GS aligns Gaussians \(\mathcal{G}\) to the underlying surface geometry. Joint Poisson fields, informed by geometric priors, then repair fragmented surfaces of both the scene and object after separation. For the object, proxy points serve as input to the Poisson fields, and the output is carved using our UNCE method to ensure geometry aligns with training observations. The Gaussians' texture properties ($\{\sigma_g,\mathcal{C}_g\}$) are refined with 2D inpainting, and this is followed by a real-time interactive simulation of the decoupled object and scene via MLS-MPM. Each stage is detailed in this section.

\subsection{3D Gaussians Preparation}
The scene is freely recorded with a consumer-level camera. The frame sequence is then processed in COLMAP \cite{schoenberger2016sfm,schoenberger2016mvs} to obtain intrinsic and extrinsic calibrations and to generate initial Gaussian centers for the next section.
\vspace{-0.4cm}
\subsubsection{Planar-based Gaussian Splatting}
\vspace{-0.1cm}
Optimizing vanilla 3D Gaussian models \cite{kerbl20233d} with only image reconstruction loss often results in local optima, complicating accurate geometry extraction, which is vital for the subsequent restoration stage. To avoid this, we adopt PGSR \cite{chen2024pgsr} for unbiased depth $D$ estimation. Given the inherent disorder of vanilla Gaussian distributions, we initially compress the Gaussians into an approximate local plane that aligns with the scene surface. This is achieved by penalizing the minimum scaling term $||\min(s_1,s_2,s_3)||_1$ during training, allowing for a tolerable loss in rendering quality to enhance geometric accuracy.  
\begin{wrapfigure}{r}{0.45\linewidth}
\vspace{-0.5em}
\hspace{-1.5em}
    \includegraphics[width=1.1\linewidth]{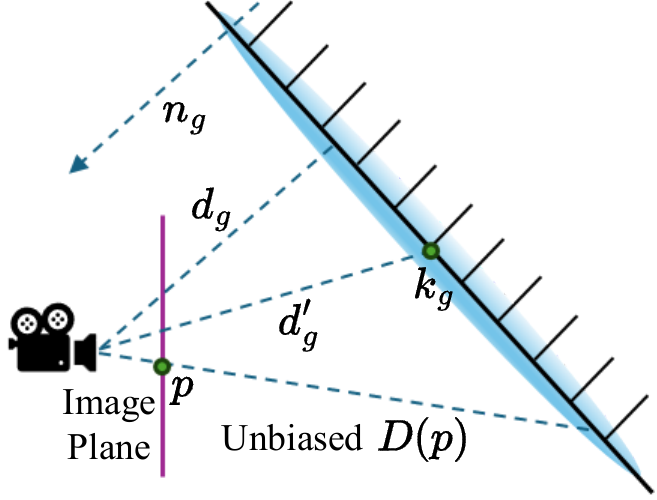}
  \vspace{-0.5em}
\end{wrapfigure}
After compression into plane-like Gaussians (see the right inset), we assign normals $\boldsymbol n_g$ along the shortest axis, with orientation disambiguated by viewing directions \cite{chen2024pgsr}. The distance to the image plane is calculated as $d_g=||\boldsymbol n_g^T \boldsymbol d'_g||$ where \( \boldsymbol{d}'_g \) is the vector from the camera center to the Gaussian center \( \boldsymbol{k}_g \). The final unbiased depth at pixel \( p \) after \(\alpha\)-blending (see \cref{eq:GS-blending}) is then given by
\vspace{-0.2cm}
\begin{equation}
\vspace{-0.2cm}
D(p) = \frac{d(p)}{\boldsymbol{n}(p) \boldsymbol{K}^{-1} p'}
\label{eq:unbiased}
\end{equation}
where \(\boldsymbol K \) is the camera's intrinsic matrix and \( p' \) is the homogeneous coordinate of \( p \).  Flattened Gaussians provide single- and multi-view geometry regularization for consistent geometry, with exposure compensation applied to address illumination variations (see \citet{chen2024pgsr} for details).
\vspace{-0.8cm}
\subsubsection{Gaussian Segmentation}
\vspace{-0.1cm}
To implement GS segmentation \cite{cen2024segment3dgaussians,gaussian_grouping}, each kernel \( g \) is assigned semantic affinity features \( \boldsymbol{\xi}_g \in \mathbb{R}^{32} \). A gating network, a single-layer MLP \( \boldsymbol{\zeta}: \mathbb{R}^{32} \to \mathbb{R}^{C} \) \cite{gaussian_grouping}, maps \(\alpha\)-blended features \( \boldsymbol{\xi}(p) \) to \( C \) segmentation class probabilities via softmax \cite{jang2022categorical}. The network is trained with cross-entropy loss using multi-view 2D segmentation labels from SAM2 \cite{ravi2024sam}. To reduce artifacts among nearby Gaussians, we apply local feature smoothing \cite{cen2024segment3dgaussians} and initialize segmentation by manually selecting classes in the first frame \cite{jiang2024vr}.

\subsection{Object-Scene Restoration}
To simulate an object \(\mathcal{O}\) interacting with its surrounding scene surface \(\mathcal{S}\), we first separate \(\mathcal{O}\) from \(\mathcal{S}\) by identifying its Gaussians through comparing affinity features with \(\alpha\)-blended \(\boldsymbol{\xi}(p)\) at a user-specified click position \(p\). We then apply KNN to remove nearby Gaussians representing residual artifacts \cite{gaussian_grouping}. For realistic simulation, we should repair and complete both \(\mathcal{O}\) and \(\mathcal{S}\), as detailed next.
\vspace{-0.4cm}
\subsubsection{Joint Poisson Fields}
\label{sec:Joint_Poisson}
\vspace{-0.1cm}
The main contribution is the novel restoration of the geometric properties \(\{\boldsymbol k_g, \boldsymbol \Sigma_g\}\) of GS, assuming that both the object \(\mathcal{O}\) and scene \(\mathcal{S}\) are smooth, closed shapes. Inspired by the equivalence between Poisson surface reconstruction and winding number field construction \cite{10.1145/3592129, feng2023winding}, we introduce \textbf{joint Poisson fields} \(\mathcal{W}\), which incorporate heterogeneous constraints to enable the simultaneous restoration of both \(\mathcal{O}\) and \(\mathcal{S}\) (see \cref{fig: joint_possion_fields}). The process is as follows:
\begin{figure}[!htb]
    \centering
    \includegraphics[width=\linewidth]{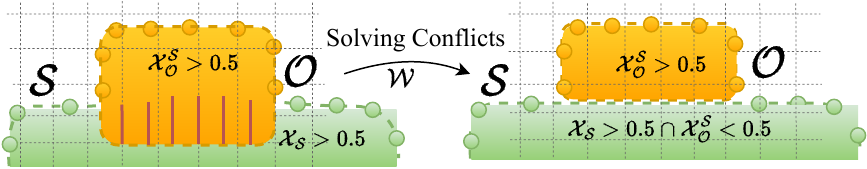}
    \vspace{-0.6cm}
    \caption{Joint Poisson Fields \(\mathcal{W}\) first reconstruct \(\mathcal{O}\) and \(\mathcal{S}\) independently, then resolve conflicts (red area) by defining a boundary that separates them into distinct, non-intersecting entities.}
    \vspace{-0.5cm}
    \label{fig: joint_possion_fields}
\end{figure}

(1) Solve the indicator functions $\mathcal{X}_{\mathcal{S}}$ and $\mathcal{X}_{\mathcal{O}}$ for the scene and object surfaces, respectively, via screened Poisson reconstruction \cite{10.1145/2487228.2487237} implicitly enforcing a minimum curvature surface. We implement this with Adaptive Multigrid Solvers \cite{kazhdan2024poisson} in corresponding canonical grid spaces, $\mathcal{W}_{\mathcal{S}}$ and $\mathcal{W}_{\mathcal{O}}$, where $\mathcal{X}>0.5$ indicates interiors while $\mathcal{X} < 0.5$ for exteriors. 
Each Poisson field with $128^3$ grid size is processed in under 20 seconds in experiments.

(2) Transform $\mathcal{X}_{\mathcal{O}}$ to $\mathcal{X}_{\mathcal{O}}^{\mathcal{S}}$ by mapping it from $\mathcal{W}_{\mathcal{O}}$ to $\mathcal{W}_{\mathcal{O}}^{\mathcal{S}}$ via world-coordinate transformation to the canonical coordinates of $\mathcal{S}$. To resolve conflicts $\{x\mid\mathcal{X}_{\mathcal{S}}(x) > 0.5 \cap \mathcal{X}_{\mathcal{O}}^{\mathcal{S}}(x) > 0.5 \}$ (intersection regions), we prioritize $\mathcal{S}$ (details in Suppl.) due to its simpler, more reliable geometry. Conflicting regions in $\mathcal{W}_{\mathcal{O}}^{\mathcal{S}}$ are then discarded.

(3) Dense interior points $P_\mathcal{O}$ (for continuum simulation) are extracted from $\mathcal{W}_{\mathcal{O}}^{\mathcal{S}}$. We apply marching cubes \cite{10.1145/37402.37422} to $\mathcal{W}_{\mathcal{S}}$ and then re-meshing \cite{pietroni2009almost} and further cropped by $P_\mathcal{O}$-scaled bounding box to get a mesh patch $\mathcal{M}_\mathcal{S}$.  Both $\mathcal{M}_\mathcal{S}$ and $P_\mathcal{O}$ are subsequently converted to world coordinates.

To solve \( \mathcal{X}_\mathcal{S} \), we use Gaussian centers $\{\boldsymbol k_g\}_{g\in\mathcal{S}}$ as input (see suppl. for normals). For \( \mathcal{X}_\mathcal{O} \), due to the geometric complexity of \( \mathcal{O} \), we introduce proxy points \( \mathcal{P}_\mathcal{O} \) as input.
\vspace{-0.4cm}
\subsubsection{Proxy Points}
\begin{figure}[!htb]
    \centering
    \vspace{-0.2cm}
    \includegraphics[width=\linewidth]{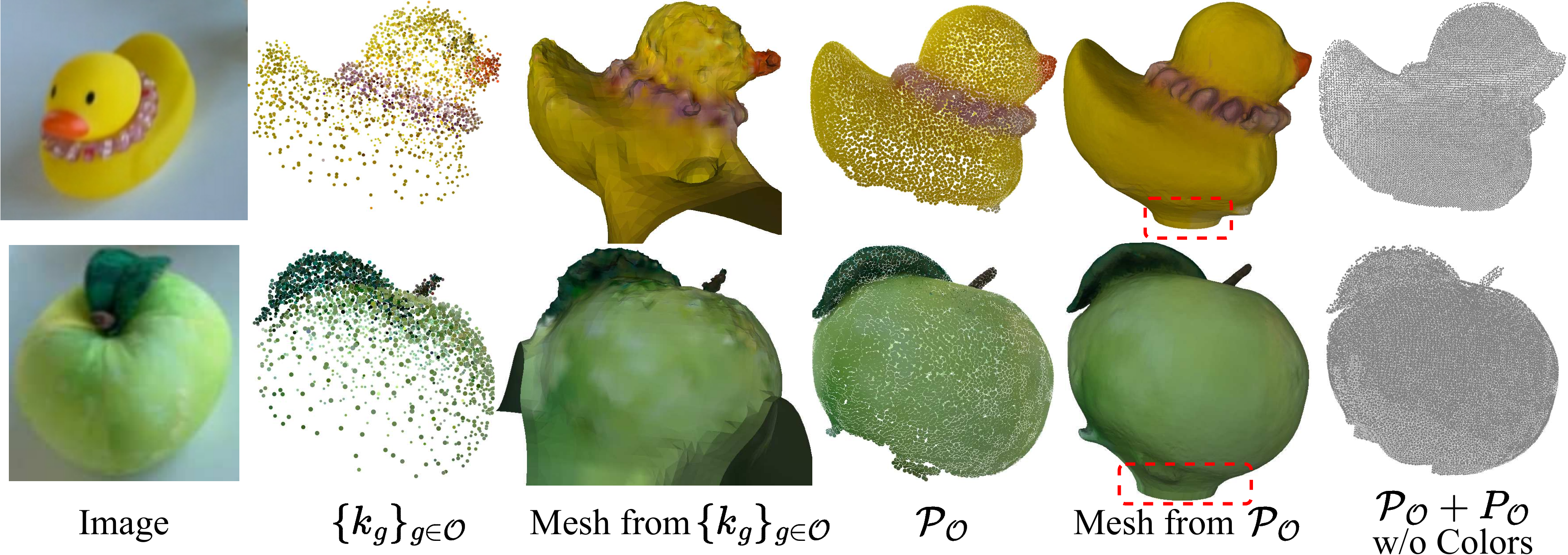}
    \vspace{-0.6cm}
    \caption{\textbf{Ablation for \(\mathcal{P}_\mathcal{O}\).} Independent Poisson reconstruction of object \(\mathcal{O}\) using Gaussian centers \(\{ \boldsymbol{k}_g \}_{g \in \mathcal{O}} \) yields poor mesh quality compared to using proxy points \(\mathcal{P}_\mathcal{O}\). Our joint Poisson field \(\mathcal{W}\), which integrates the scene surface \(\mathcal{S}\), effectively removes the overextended regions (highlighted in red). The final dense points \(P_\mathcal{O}\) are then combined with proxy points \(\mathcal{P}_\mathcal{O}\) for Gaussian restoration and continuum simulation.}
    \vspace{-0.4cm}
    \label{fig: ablation_for_proxy}
\end{figure}

\vspace{-0.1cm}
Due to \(\alpha\)-blending, Gaussian centers $\{\boldsymbol k_g\}_{g\in\mathcal{O}}$ fail to accurately represent the complex surface of \(\mathcal{O}\). Our proposed proxy points \(\mathcal{P}_\mathcal{O}\) can enhance geometry estimations of $\mathcal{X}_\mathcal{O}$ ablated as shown in \cref{fig: ablation_for_proxy}. 

We first render RGB images and unbiased depth maps \(D\)  in \cref{eq:unbiased} for the entire layout under all training views. Next, we obtain the projected mask \(M_\mathcal{O}^{\text{proj}}\) by setting zero-opacity for all other Gaussians \((\mathcal{G} \setminus \mathcal{O})\), where zero in \(M_\mathcal{O}^{\text{proj}}\) indicates no accumulated opacity, while one signifies existing opacity at a pixel location. Using the masked projected depth map \(D \circ M_\mathcal{O}^{\text{proj}}\), the TSDF fusion algorithm \cite{newcombe2011kinectfusion} concentrates on the object area, rapidly integrating information from training views within 100 seconds, followed by standard post-processing \cite{chen2024pgsr}. However, the integrated result still includes points from \((\mathcal{G} \setminus \mathcal{O})\) due to boundaries smearing \cite{wang2024self} from \(M_\mathcal{O}^{\text{proj}}\). To address this, we segment the final proxy points \(\mathcal{P}_\mathcal{O}\) by inheriting features \(\boldsymbol{\xi}_g\) from raw Gaussian kernels of \(\mathcal{O}\) with nearest neighbor search.
\vspace{-0.4cm}
\subsubsection{Unilateral Negative Cross Entropy}
\vspace{-0.1cm}
Despite closing broken surfaces and filling internal dense points \( \mathcal{P}_\mathcal{O} \), the over-smoothness of Poisson fields leads to \emph{geometry expansion}, introducing particles beyond observable viewpoints. To address this, we apply \emph{multi-view carving}. Specifically, we propose a Unilateral Negative Cross Entropy (UNCE) loss at each rendered pixel \( p \) for the \textbf{isometric} dense object Gaussians \( G_\mathcal{O} \). This loss measures the discrepancy between the \(\alpha\)-blended opacity \( \mathds{1}_\mathcal{O} \) (see \cref{eq:GS-blending}) during fine-tuning and the 2D ground truth object mask \( M_\mathcal{O}^{\text{GT}} \) from SAM2, defined as:
\begin{equation}
    \text{UNCE}(p) = -(1-M_\mathcal{O}^{\text{GT}}(p))\log(1-\mathds{1}_\mathcal{O}(p)).
\end{equation}
Every 100 iterations, we clean Gaussians $\{\sigma_g \leq 0.05\}$. These isometric Gaussians \( G_\mathcal{O} \) are defined by centers \(\{\boldsymbol{k}_g \in \mathcal{P}_\mathcal{O} \cup P_\mathcal{O}\} \), each associated with an opacity of \( \sigma_g = 0.1 \) and an isometric covariance matrix \( \boldsymbol \Sigma_g = \text{diag}(s_g^2, s_g^2, s_g^2) \) \cite{cai2024gaussian}. Here, \( s_g = c\left(\frac{3}{4\pi}\right)^{\frac{1}{3}} \) \cite{xie2024physgaussian}, where \( c \) is the Poisson grid cell length in world coordinates. For centers \( \{\boldsymbol k_g \in \mathcal{P}_\mathcal{O}\} \), view-independent SHs are derived from the colors of points already integrated by the TSDF.  In contrast, view-independent SHs for \( \{\boldsymbol k_g \in P_\mathcal{O}\} \) are Gaussian-weighted interpolations based on the 15 nearest neighbors in \( \mathcal{P}_\mathcal{O} \). We zero all other coefficients in each \( \mathcal{C}_g \).

\vspace{-4mm}
\subsubsection{Gaussian Restoration}
\vspace{-0.1cm}
During multi-view carving, we also fine-tune \( \{\sigma_g,\mathcal{C}_g\}_{g \in G_\mathcal{O}} \) as described by \citet{lipac}, but using \( M_\mathcal{O}^{\text{GT}} \)-masked training images to mitigate influence from other areas in the scene. In each iteration, a random background is applied for image rendering and is consistently used for the other regions of the masked ground truth. 

To restore the Gaussians (i.e., the holes) in the scene surface $\mathcal{S}$, we first bind new 3D flattened Gaussians $G_\mathcal{S}$ \cite{guedon2024sugar,borycki2024gasp} to the patch mesh $\mathcal{M}_S$ (see our Mesh to Gaussian algorithm in Supplementary) with minimal scaling $\epsilon$ along the mesh face normals. At this stage, we finalize and fix the geometric properties $\{\boldsymbol k_g, \boldsymbol \Sigma_g\}_{g \in G_\mathcal{S}}$. During fine-tuning, we adjust only the texture properties $\{\sigma_g, \mathcal{C}_g\}_{g \in G_\mathcal{S}}$, initialized from the nearest neighbors of the raw broken $\mathcal{S}$, guided by 2D inpainted images in the masked areas $M_\mathcal{O}^{\text{GT}}$ using LaMa \cite{suvorov2022resolution}. Finally, we fill holes in $\mathcal{S}$ by adding the patch $G_\mathcal{S}$.
\subsection{Interactive Simulation}
We simulate and render all restored Gaussians \( G_\mathcal{O} \) to enable a range of interactive simulations, including user-specified impulses as external forces for elastic deformation, scene collisions with \( \mathcal{S} \), and effects like shape fracturing and material changes, all based on MLS-MPM. To enforce a \emph{Dirichlet boundary condition} \cite{bazilevs2007weak}, we set the velocities of grid nodes containing Gaussians from the restored scene \( \mathcal{S} \) to zero during Grid Operation stage in MLS-MPM, creating a sticky boundary effect. To simulate gravity, we automatically align the z-axis by segmenting Gaussians for planar objects (e.g., ground or desk surfaces) and estimating the plane normals with RANSAC \cite{li2017improved}. We then apply the rotation matrix derived from plane normals to all \( \{\boldsymbol{k}_g, \boldsymbol{\Sigma}_g\}_{g \in \mathcal{G}} \) directly, while view-dependent SHs are rotated through Wigner D-matrices \cite{wigner2012group} (see details in Suppl.)

\section{Experiments}
\subsection{Implementation Details}
Input resolutions range from 720p to 1K. Gaussian restoration of \( \mathcal{S} \) and \( \mathcal{O} \) uses \( L_1 \) and \( L_{\text{SSIM}} \) losses with UNCE regularization at \( 10^{-4} \), fine-tuned for 1000 iterations for \( \mathcal{S} \) and 3000 for \( \mathcal{O} \), skipping iterations without valid masks, totaling under 4 minutes. For LaMa inpainting, masked areas are dilated with a \( 21 \times 21 \) kernel to reduce boundary artifacts in \( M_{\mathcal{O}}^{\text{GT}}\) \cite{liu2024infusion}. The simulation area and physical parameters (e.g., \( E, \mu \)) are manually set following \cite{xie2024physgaussian, jiang2024vr} (see Supplementary). Based on Warp \cite{warp2022}, the simulation runs on an 18-core Intel Xeon Gold 5220 CPU and NVIDIA A40 GPU, achieving \(\sim\)10 FPS for 50-frame videos.

\subsection{Evaluating Object-Scene Interaction}
\begin{figure*}[!htb]
    \centering
    \includegraphics[width=\linewidth]{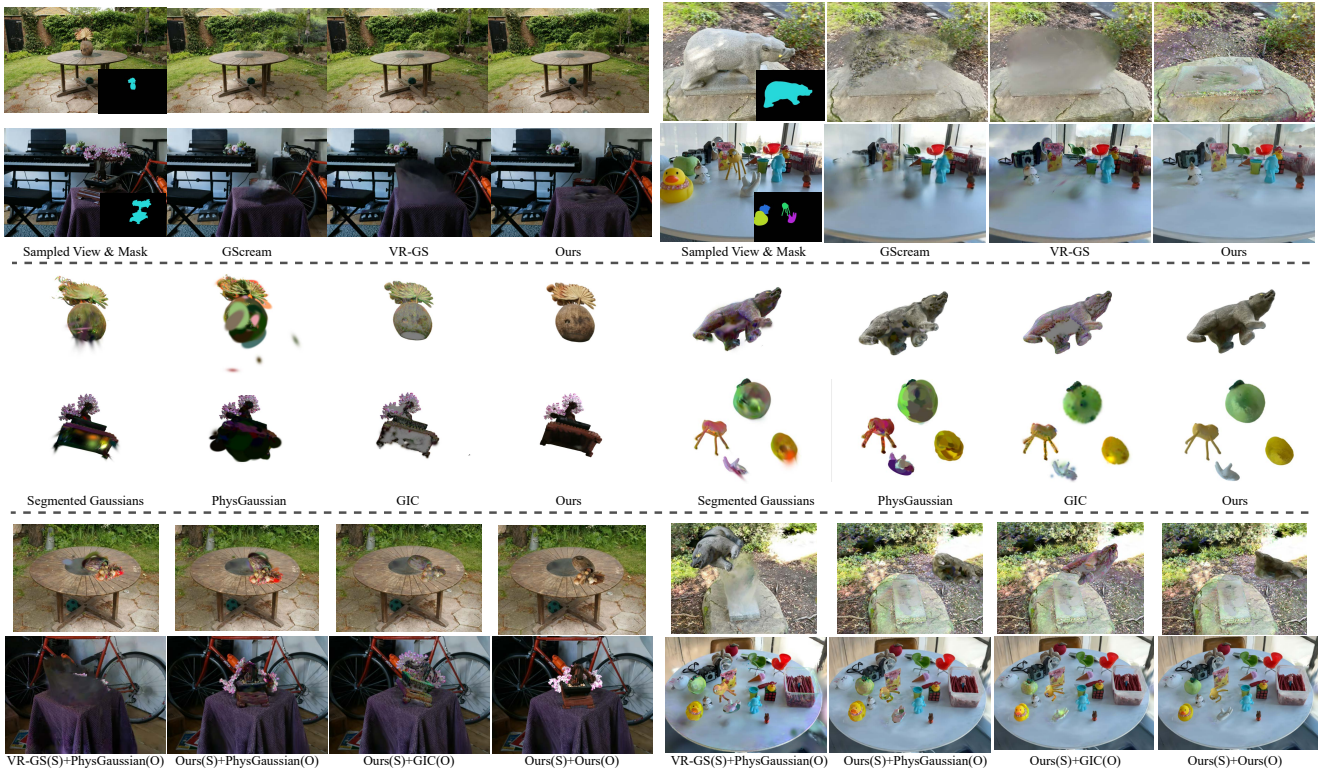}
    \vspace{-0.6cm}
\caption{\textbf{Qualitative Comparisons.} We demonstrate \textbf{Scene Restoration} (Top), \textbf{Object Restoration} (Middle), and \textbf{Object-Scene Interactive Simulation} (Bottom) on real-world scenes, including GARDEN, BEAR, BONSAI, and FIGURINES. For each comparison, a single test viewpoint is selected, with the object initially suspended before being impacted by an external force during simulation.}
    \label{fig: Qualitative_Comparision}
\vspace{-5mm}
\end{figure*}
\paragraph{\emph{Dataset}} We evaluate our system for generating diverse object-scene interactive simulations using several real-world scenario sources. Our evaluation includes the following datasets: BICYCLE, GARDEN, BONSAI, ROOM, and KITCHEN from the Mip-NeRF360 \cite{barron2022mip} dataset; TRUCK from the Tanks\&Temples dataset \cite{knapitsch2017tanks}; PLAYROOM from the Deep Blending dataset \cite{hedman2018deep}; FIGURINES from Lerf \cite{kerr2023lerf}; and BEAR from Instruct-NeRF2NeRF \cite{haque2023instruct}.
\vspace{-4mm}
\paragraph{\emph{Baselines}} We compare our method with SOTA simulation frameworks based on Gaussian splatting, 
 incorporating necessary adaptations: 1) \textbf{PhysGaussian} \cite{xie2024physgaussian} uses anisotropy regularization to prevent narrow kernels and applies a user-defined opacity field (based on \( \mathcal{O} \)’s bounding box) for interior filling. 2) \textbf{GIC} \cite{cai2024gaussian} employs isotropic Gaussians with a coarse-to-fine density field to fill interior points, assigning scales and opacities with zero color. 3) \textbf{VR-GS} \cite{jiang2024vr} is closely aligned with our approach; however, due to unavailable simulation code, we adapt their methods to restore scene Gaussians, \( \mathcal{S} \), with geometry and texture properties guided by LaMa. 4) \textbf{GScream} \cite{wang2024gscream}, a SOTA technique for \( \mathcal{S} \) restoration, integrates monocular depth estimation \cite{ke2023repurposing} from the inpainted reference view for training guidance.
\vspace{-4mm}
\paragraph{\emph{User Study}} We conducted a human evaluation to assess both visual realism and simulation fidelity,  following methods from prior work \cite{liu2025physgen,chen2023videocrafter1,wei2024dreamvideo}. Ten participants with varying experience in simulation and 3D vision rated three aspects: 1) \textbf{Scene Restoration Quality (SRQ)}, which evaluates the accuracy of scene restoration, \( \mathcal{S} \), after object removal; 2) \textbf{Object Restoration Quality (ORQ)}, assessing the realism of restored objects, \( \mathcal{O} \); and 3) \textbf{Interactive Simulation Fidelity (ISF)}, checking if the object scenes response to a user-specified impulse is both realistic and as expected. Rendered videos of \( \mathcal{S} \), \( \mathcal{O} \), and interactive simulations were presented in random order, with participants rating each on a five-point scale (1 = poor, 5 = excellent). Mean scores are reported, with supplementary material containing additional statistics and video examples.
\definecolor{LightCyan}{rgb}{0.88,1,1}
\begin{table}[!htb]
\centering
\vspace{-2mm}
\caption{\textbf{User Study}. Participants rated the fidelity of restoration and interactive simulation in a moving-camera video.}\label{tab: human_evaluation}
\vspace{-2mm}
\resizebox{\linewidth}{!}{
\begin{tabular}{ccc|cc}
\toprule
\multicolumn{3}{c}{\textbf{Scene Restoration}} & \multicolumn{2}{c}{\textbf{Object Restoration}} \\
Methods & SRQ $\uparrow$ & Time $\downarrow$ & Methods & ORQ $\uparrow$ \\
\midrule
GScream \cite{wang2024gscream} & 1.94  & $\sim$70m & PhysGaussian \cite{xie2024physgaussian} & 1.40 \\
VR-GS \cite{jiang2024vr} & 2.12  & $\sim$7m  & GIC \cite{cai2024gaussian}  & 1.60 \\
\cellcolor{LightCyan}\textbf{Ours} & \cellcolor{LightCyan}\textbf{3.48} & \cellcolor{LightCyan}\textbf{$\sim$1m} & \cellcolor{LightCyan}\textbf{Ours} & \cellcolor{LightCyan}\textbf{4.03} \\
\midrule
\multicolumn{5}{c}{\textbf{Object-Scene Interactive Simulation}} \\
\multicolumn{3}{c}{Methods}& \multicolumn{2}{c}{ISF $\uparrow$} \\
\midrule
\multicolumn{3}{c}{VR-GS($\mathcal{S}$) + PhysGaussian($\mathcal{O}$)}  & \multicolumn{2}{c}{1.50} \\
\multicolumn{3}{c}{Ours($\mathcal{S}$) + PhysGaussian($\mathcal{O}$)} & \multicolumn{2}{c}{2.60}  \\
\multicolumn{3}{c}{Ours($\mathcal{S}$) + GIC($\mathcal{O}$)} & \multicolumn{2}{c}{2.73} \\
\multicolumn{3}{c}{\cellcolor{LightCyan}\textbf{Ours($\mathcal{S}$) + Ours($\mathcal{O}$)}} & \multicolumn{2}{c}{\cellcolor{LightCyan}\textbf{4.35}} \\
\bottomrule
\vspace{-6mm}
\end{tabular}}
\end{table}

\vspace{-4mm}
\begin{figure*}
    \centering
    \includegraphics[width=\linewidth]{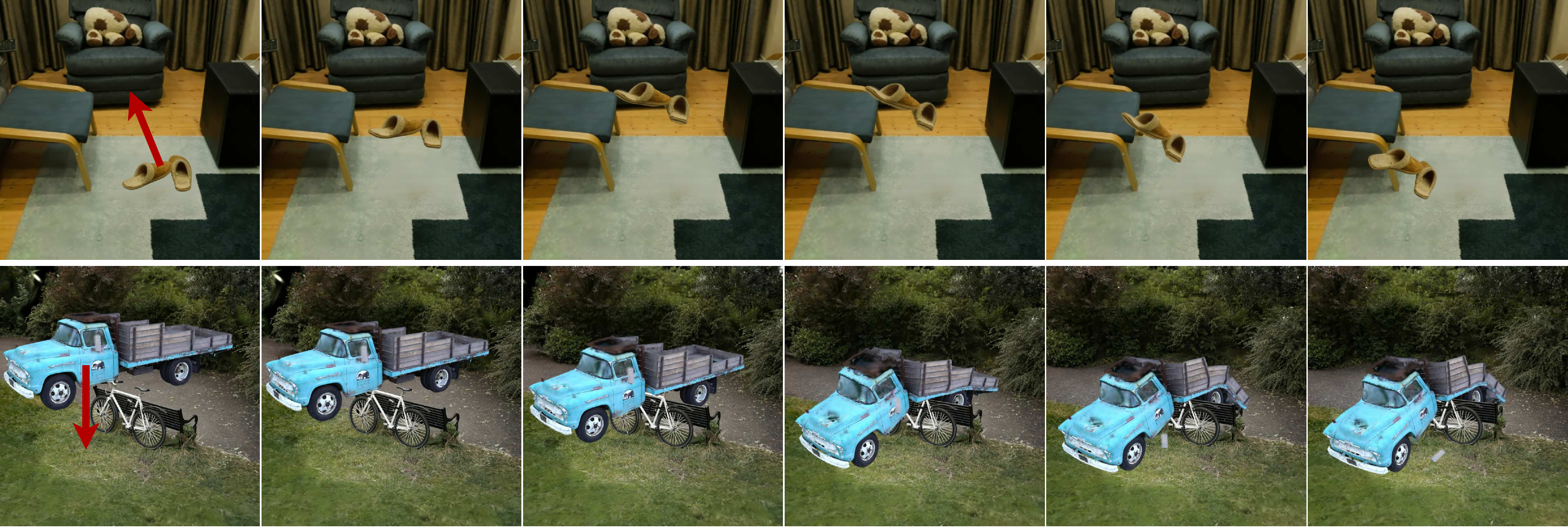}
    \vspace{-0.7cm}
    \caption{\textbf{Interactive Versatility.} Our object-scene decoupling method enables a variety of user-specified interactions both within a single scene (e.g., ROOM Top) and across different scenes (e.g., TRUCK in BICYCLE, Bottom).
    }
    \label{fig: various_material}
    \vspace{-4mm}
\end{figure*}
\paragraph{\emph{Results}} Our method (\cref{tab: human_evaluation}) achieves the highest ratings and shortest training time (1 minute) for scene restoration. VR-GS and GScream rely on 2D inpainting for \( \mathcal{S} \) restoration, leading to geometry inaccuracies (e.g., BEAR, BONSAI in \cref{fig: Qualitative_Comparision} Top) when inpainting quality is poor. GScream’s use of a single reference image limits view consistency, causing issues in non-forward-facing views. In contrast, our approach uses planar-based GS geometry priors, ensuring precise structural restoration while limiting 2D inpainting to texture properties. For \( \mathcal{O} \) restoration, we are the first to restore \( \mathcal{O} \) in both single and complex multi-object scenes (e.g., FIGURINES in \cref{fig: Qualitative_Comparision} Middle), maintaining input quality. Unlike PhysGaussian, which suffers from artifacts due to incomplete opacity assumptions, and GIC, which shows artifacts from non-zero internal opacities (white dots in \cref{fig: Qualitative_Comparision} Middle), our method produces stable, high-quality results. For interactive simulations, GIC (BONSAI in \cref{fig: Qualitative_Comparision} Bottom) exhibits unintended motion due to particle imbalance. VR-GS, relying on 2D inpainting, shows flawed geometry of \( \mathcal{S} \), limiting object-scene interactions and causing artifacts or pass-through issues (e.g., BEAR, BONSAI in \cref{fig: Qualitative_Comparision} Bottom). Our video demonstrates dynamic effects, and \cref{fig: various_material} showcases simulations with user-specified impulses, including cross-scene interactions (e.g., TRUCK in BICYCLE scene), highlighting our method’s high controllability and motion realism.

\subsection{Decoupling Benchmark Evaluation}
\begin{figure}
    \centering
    \includegraphics[width=\linewidth]{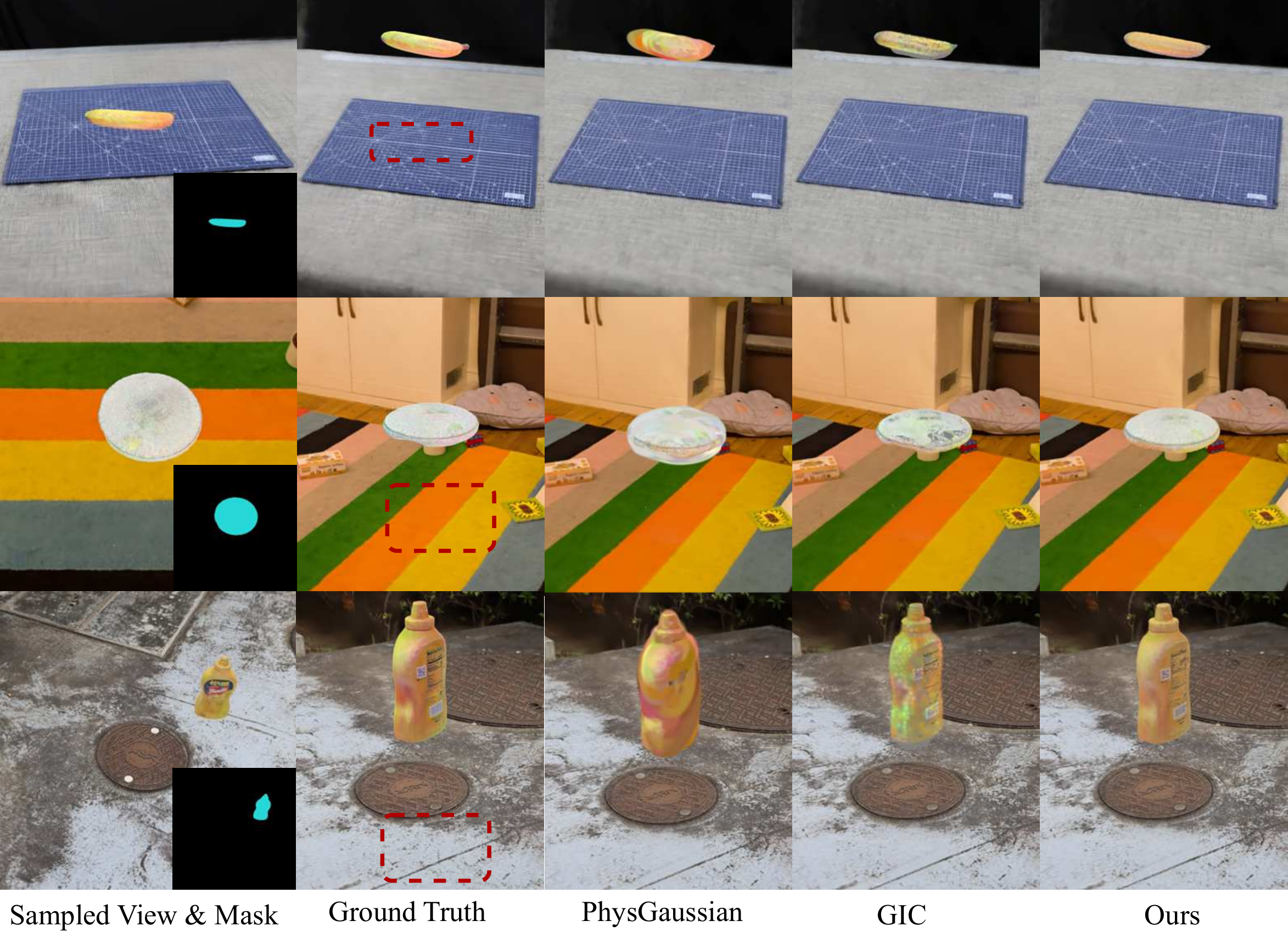}
    \vspace{-4mm}
    \caption{\textbf{Benchmark Comparisons.} A test viewpoint visualizes comparisons using the restored scene from our method, with inpainting regions marked by a red rectangle in the Ground Truth.}
    \label{fig: benchamakr_qualitative}
    \vspace{-5mm}
\end{figure}
\paragraph{\emph{Dataset}} To address the lack of ground truth for object-scene decoupling interactions, we utilize real reconstructed scenes and objects from the PEGASET dataset \cite{PEGASUS2024} and the PLAYROOM and SOFA SUITE environments from BlenderNeRF \cite{chen2024blender}. Test cases with realistic elements are created by placing objects within scenes using PyBullet \cite{coumans2016pybullet} and rendering object-scene setups as input from the raw scene’s training viewpoints. Ground truth for object restoration is provided by well-reconstructed objects, internally filled  \cite{xie2024physgaussian}, and without internal textures, while ground truth for scene restoration is based on the raw scenes with no objects. For object-scene interaction, we render multi-view MLS-MPM simulations by dropping these ground truth objects from a height to the ground, following \cite{cai2024gaussian}.
\vspace{-4mm}
\paragraph{\emph{Metrics}} 
We use \textbf{PSNR} \cite{hore2010image}, \textbf{LPIPS} \cite{zhang2018unreasonable}, and \textbf{FID} \cite{heusel2017gans} as primary metrics to evaluate reconstruction quality. For restoring scene \( \mathcal{S} \) and object \( \mathcal{O} \), we additionally apply Chamfer Distance (\textbf{CD}) \cite{fan2017point} to measure the geometric accuracy of Gaussian centers within inpainted regions, critical for accurate physics-based simulation. Viewpoints outside training views are used for \( \mathcal{O} \) captures while training viewpoints are retained for \( \mathcal{S} \). To evaluate motion accuracy in interactive simulations, we compute \textbf{Motion-FID} \cite{liu2025physgen} by extracting and colorizing optical flow using RAFT \cite{teed2020raft} and calculating FID on the resulting flow images.
\vspace{-4mm}
\begin{table}[!htb]
\centering
\caption{\textbf{Quantitative Comparisons \& Ablations.} We create a decoupling benchmark with comprehensive metrics comparing baselines and ablations to validate design choices.}\label{tab: quantitative_comparisions}
\vspace{-1mm}
\resizebox{\linewidth}{!}{
\begin{tabular}{ccccc}
\toprule
\multicolumn{5}{c}{\textbf{Scene Restoration}}\\
Methods & PSNR $\uparrow$ & LPIPS $\downarrow$ & \multicolumn{1}{c}{FID $\downarrow$}& \multicolumn{1}{c}{CD ($\times 10^{-3}$) $\downarrow$}\\
\midrule
GScream \cite{wang2024gscream} & 17.82  & 0.56 & 42.28 & 44.00 \\
VR-GS \cite{jiang2024vr} & 25.13  & 0.32 & 58.50 & 6.41 \\
\cellcolor{LightCyan}\textbf{Ours} & \cellcolor{LightCyan}\textbf{27.32}  & \cellcolor{LightCyan}\textbf{0.30} & \cellcolor{LightCyan}\textbf{32.07} &\cellcolor{LightCyan}\textbf{4.40}\\
\midrule
\multicolumn{5}{c}{\textbf{Object Restoration}}\\
Methods & PSNR $\uparrow$ & LPIPS $\downarrow$ & \multicolumn{1}{c}{FID $\downarrow$}& \multicolumn{1}{c}{CD ($\times 10^{-3}$) $\downarrow$}\\
\midrule
 PhysGaussian \cite{xie2024physgaussian} &24.46& 0.07 & 227.60 & 0.53 \\
GIC \cite{cai2024gaussian} & 26.62& 0.06 & 201.91 & 0.73\\
\cellcolor{LightCyan}\textbf{Ours} & \cellcolor{LightCyan}\textbf{30.32}& \cellcolor{LightCyan}\textbf{0.04} & \cellcolor{LightCyan}\textbf{138.75} &\cellcolor{LightCyan}\textbf{0.17}\\
\midrule
\multicolumn{5}{c}{\textbf{Object-Scene Interaction Simulation}}\\
Methods & PSNR $\uparrow$ & LPIPS $\downarrow$ & \multicolumn{1}{c}{FID $\downarrow$}& \multicolumn{1}{c}{Motion-FID $\downarrow$}\\
\midrule
 PhysGaussian \cite{xie2024physgaussian} &19.48& 0.37 & 112.55 & 54.79 \\
GIC \cite{cai2024gaussian} & 20.90& 0.31 & 134.56 & 47.47\\
\midrule
 w/o dense $P_\mathcal{O}$&21.19& 0.29 & 98.19 & 48.39 \\
  w/o Proxy $\mathcal{P}_\mathcal{O}$&21.08& 0.30 & 90.26 & 36.01 \\
   w/o $\mathcal{W}$ &20.97& 0.30 & 96.16 & 42.27 \\
\cellcolor{LightCyan}\textbf{Ours} & \cellcolor{LightCyan}\textbf{21.33}& \cellcolor{LightCyan}\textbf{0.29} & \cellcolor{LightCyan}\textbf{86.98} &\cellcolor{LightCyan}\textbf{31.69}\\
\bottomrule
\end{tabular}}
\vspace{-5mm}
\end{table}
\paragraph{\emph{Results}}For object-scene interaction, we use our restored \( \mathcal{S} \) across all methods to ensure fair comparison. Quantitative results are shown in \cref{tab: quantitative_comparisions} and qualitative examples in \cref{fig: benchamakr_qualitative}, with sample views of objects attached to scene surfaces from the input. Our restoration of \( \mathcal{S} \) closely matches ground truth, and \( \mathcal{O} \) significantly outperforms other methods in GS simulation. Competing methods often produce artifacts due to inadequate handling of broken surfaces and hidden areas, which degrades interactive simulation quality. Although our approach excels in interactive simulation, the object's FID is high due to reliance on training view interpolation for texture restoration. Future work will explore 3D AI-based texture generative inpainting to improve this.
\begin{figure}[!htb]
    \centering
    \vspace{-2mm}
    \includegraphics[width=\linewidth]{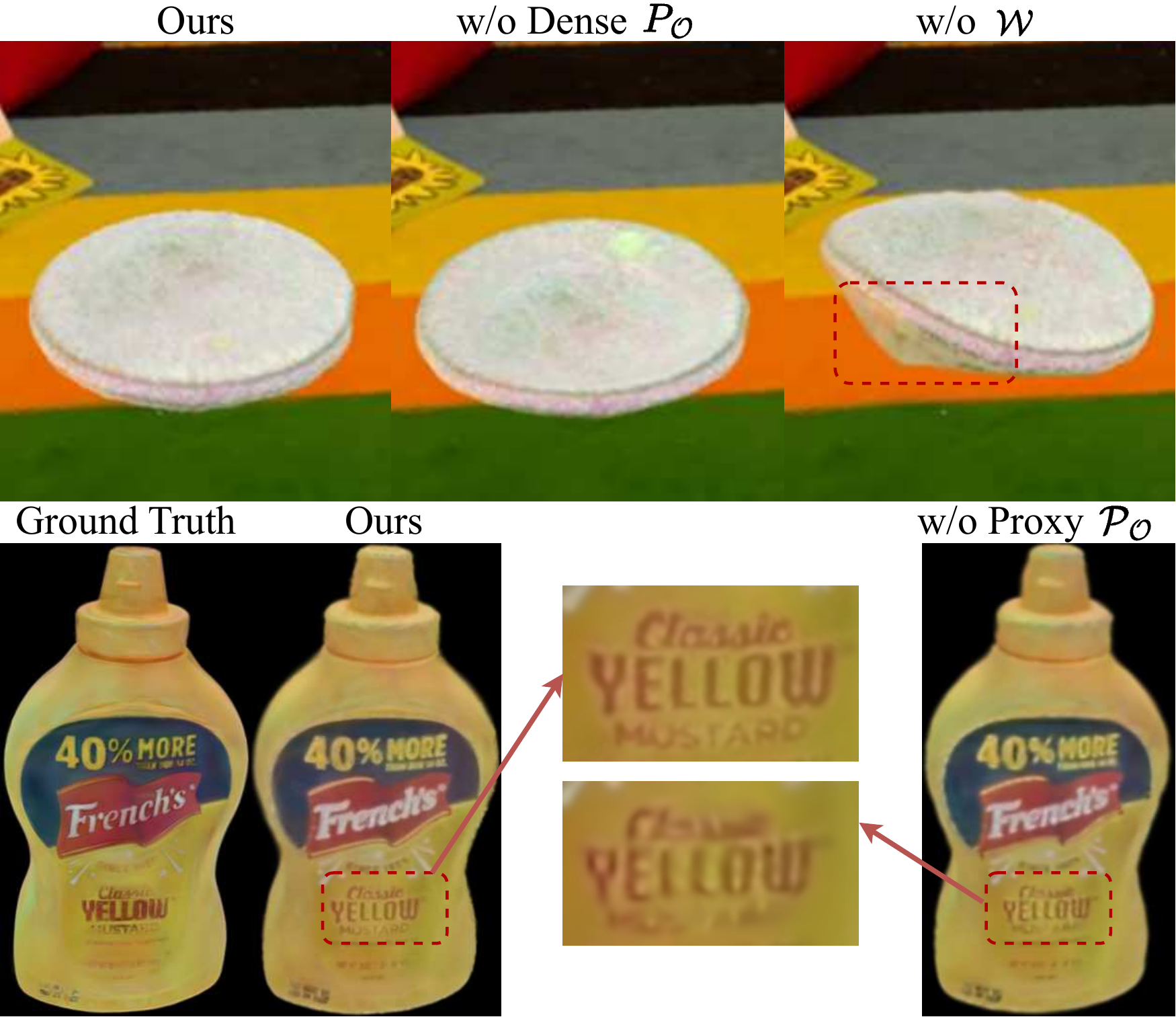}
    \caption{\textbf{Dense $P_\mathcal{O}$} prevents collapse under gravity, \textbf{Joint Point Fields $\mathcal{W}$} remove red-highlighted intersection regions, and \textbf{Proxy Points $\mathcal{P}_\mathcal{O}$} enhance texture details (see zoom-ins).}
    \label{fig: qualitative_ablation}
\vspace{-6mm}
\end{figure}
\vspace{-1mm}
\paragraph{\emph{Ablations}} We quantitatively evaluate several design choices (see \cref{tab: quantitative_comparisions}): 1) \textbf{Dense Interior Points ($P_\mathcal{O}$)}, our internal filling strategy, prevent collapse under gravity or external forces, unlike objects without internal particles (see \cref{fig: qualitative_ablation}, Top-Middle). 2) \textbf{Proxy Points ($\mathcal{P}_\mathcal{O}$)} enhance geometry recovery in Poisson reconstruction (see \cref{fig: ablation_for_proxy}), and their combination with $P_\mathcal{O}$ improves texture details over $P_\mathcal{O}$ alone (see \cref{fig: qualitative_ablation}, Bottom). 3) \textbf{Joint Poisson Fields ($\mathcal{W}$)} reduce artifacts and resolve intersection regions better than independent Poisson reconstructions (see \cref{fig: qualitative_ablation}, Top-Right).
\subsection{Additional Qualitative Ablations}
\begin{figure}[!htb]
    \centering
    \vspace{-6mm}
    \includegraphics[width=\linewidth]{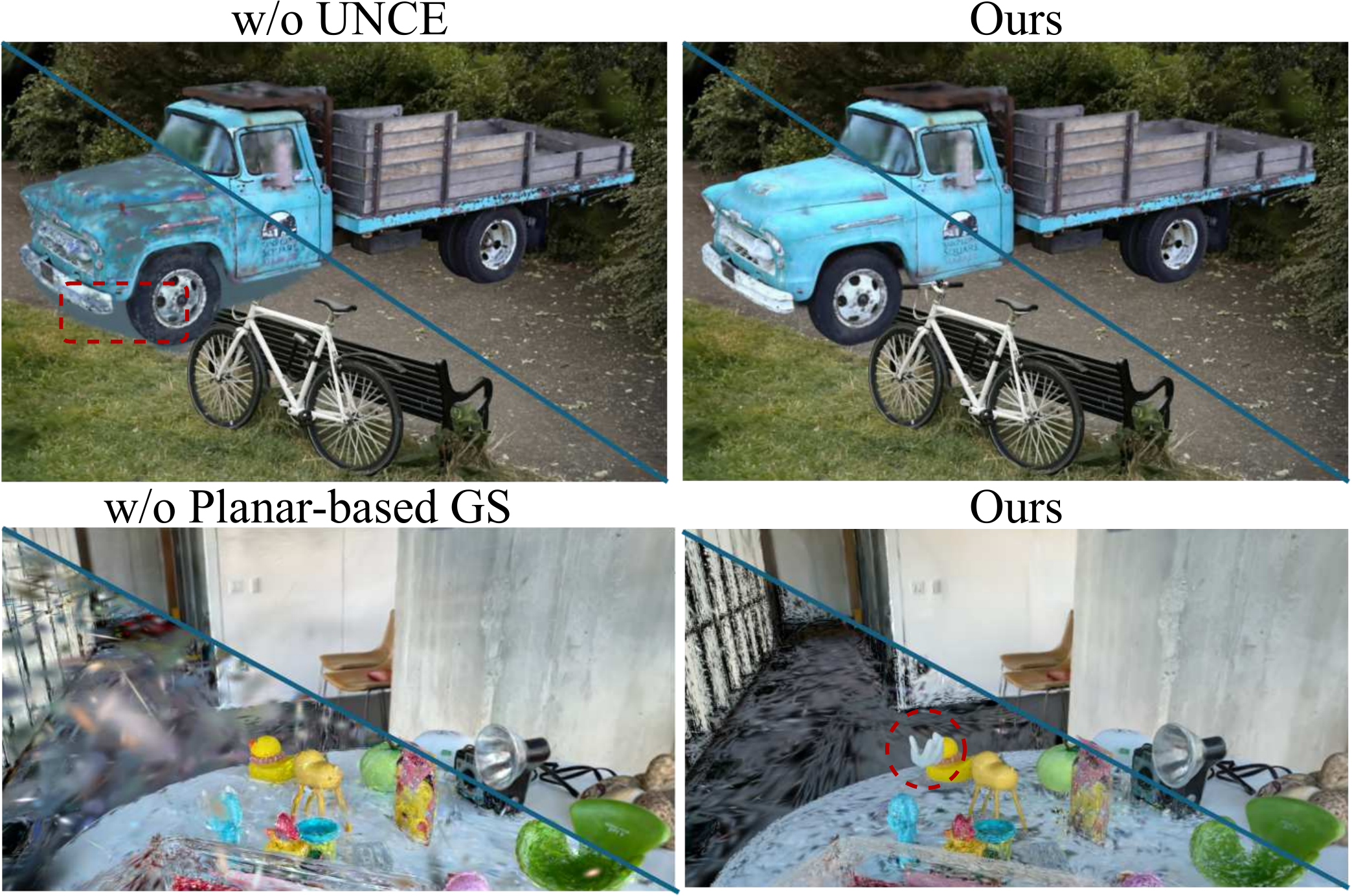}
    \vspace{-0.7cm}
    \caption{\textbf{UNCE} (Top) removes artifacts from Poisson expansion via multi-view carving. Opacity is set to one for TRUCK to highlight artifacts. \textbf{Planar-based GS} (Bottom) avoids floaters and artifacts compared to Vanilla GS \cite{kerbl20233d}, which limits motion (e.g., red-circled figurine). Opacity is set to 1, with $\times 0.4$ scaling for better Gaussian kernel visualization.}
    \label{fig: planr_based_ablation}
    \vspace{-0.7cm}
\end{figure}

\paragraph{\emph{UNCE}} Poisson reconstruction prioritizing smoothness in \( \mathcal{W} \) can introduce artifacts, even with opacity filtering. Our Unilateral Negative Cross Entropy (UNCE) method (shown in \cref{fig: planr_based_ablation}, Top) leverages negative labels from SAM2 to carve and remove these artifacts, aligning \( G_\mathcal{O} \) with the underlying geometry for accurate simulation.
\vspace{-4mm}
\paragraph{\emph{Planar-based GS}} Unlike standard Gaussian splatting \cite{kerbl20233d} with low-opacity filtering (\(\sigma_g \leq 0.02\)) \cite{xie2024physgaussian}, planar-based GS with compressed kernels enhances geometry regularization, reducing floaters (\cref{fig: planr_based_ablation}, Bottom) without the need for opacity filtering. This method enables unrestricted object motion in the simulation area, free from scene artifacts.
\section{Discussion}

\paragraph{Conclusion} This paper presents \METHOD, a fast and robust approach for decoupling static objects from contact surfaces and restoring geometry and texture for object-scene interaction using the MLS-MPM simulator.  
\vspace{-4mm}
\paragraph{Limitations} Our evaluation does not address complex scenes with multiple objects in varying contact configurations. High-frequency texture completion for object restoration is challenging, and GS-based texture generative approaches \cite{yu2024instantstylegaussian, jain2024stylesplat, zhang2024stylizedgs} may offer potential solutions. Additionally, decoupling the fine-grained components \cite{yang2024sampart3d} of individual objects presents further difficulties.

\vspace{-8pt}

\vspace{-1em}

\newpage
{
    \small
    \bibliographystyle{ieeenat_fullname}
    \bibliography{main}
}


\end{document}


\definecolor{cvprblue}{rgb}{0.21,0.49,0.74}
\maketitlesupplementary

\appendix
\setcounter{table}{0}
\renewcommand{\thetable}{A\arabic{table}}
\setcounter{figure}{0}
\renewcommand{\thefigure}{A\arabic{figure}}
\section{Organization}
In this paper, we introduce \textbf{DecoupledGaussian}, a fast and robust method for decoupling static objects from contact surfaces while restoring geometry and texture for improved object-scene interaction. Using the MLS-MPM simulator, our approach extends beyond rigid 3D reconstructions, enabling more dynamic and flexible applications of Gaussian Scene (GS) representations. We encourage readers to view the accompanying video for demos of the dynamic effects. This supplementary material provides detailed material parameters, methodology, and additional experiments to offer a comprehensive understanding of our approach.

\textbf{Note:} Figures, sections, and tables in the supplementary material are prefixed with a letter for distinction, while those without a prefix refer to content in the main paper.
\section{Material Properties}  
We use two constitutive models from \citet{10.1145/3610548.3618207}: Fixed Corotated and Drucker-Prager Plasticity. The material parameters, including Young's modulus ($E$) and shear modulus ($\mu$) (Sec. 5.1), for each case are summarized in Table \ref{tab:material_properties}. 
\begin{table}[!htb]
    \centering
\caption{\textbf{Material Properties Configuration.}}
    \resizebox{\linewidth}{!}{
    \begin{tabular}{c c c c c}
    \toprule
         Case & Figure & Constitutive Model & $\mu$ & $E$ \\
    \midrule
      \texttt{Bear\_collisions}   & Fig. 1 & Fixed Corotated            & 0.3 & $3 \times 10^6$ \\
      \texttt{Bear\_melting}      & Fig. 1 & Drucker-Prager Plasticity  & 0.3 & $3 \times 10^6$ \\
      \texttt{Kitchen}            & Fig. 3 & Fixed Corotated            & 0.3 & $3 \times 10^6$ \\
      \texttt{Garden\_collisions} & Fig. 6 & Fixed Corotated            & 0.3 & $3 \times 10^6$ \\
      \texttt{Bonsai\_collisions} & Fig. 6 & Fixed Corotated            & 0.4 & $2 \times 10^6$ \\
      \texttt{Figurines\_collisions} & Fig. 6 & Fixed Corotated          & 0.3 & $3 \times 10^6$ \\
      \texttt{Room}               & Fig. 7 & Fixed Corotated            & 0.3 & $3 \times 10^6$ \\
      \texttt{Truck\_Bicycle}     & Fig. 7 & Fixed Corotated            & 0.3 & $3 \times 10^6$ \\
      \texttt{Banana}             & Fig. 8 & Fixed Corotated            & 0.3 & $3 \times 10^6$ \\
      \texttt{Pillow}             & Fig. 8 & Fixed Corotated            & 0.3 & $3 \times 10^6$ \\
      \texttt{Mustard}            & Fig. 8 & Fixed Corotated            & 0.3 & $3 \times 10^6$ \\
        \texttt{Bonsai}            & \cref{fig: simulation_qualitative} & Drucker-Prager Plasticity          & 0.4 & $2 \times 10^6$ \\
        \texttt{Kitchen}            & \cref{fig: simulation_qualitative} & Drucker-Prager Plasticity          & 0.3 & $3 \times 10^6$ \\    
    \bottomrule
    \end{tabular}}
    \label{tab:material_properties}
\end{table}

\section{Technique Details}
\subsection{Wigner D-matrix}
The Wigner D-matrix \cite{wigner1931gruppentheorie,shen2018approximations} (Sec. 4.3) \( D_{m', m}^{(j)}(\alpha, \beta, \gamma) \) describes the rotation of a function on the sphere in terms of Euler angles \( (\alpha, \beta, \gamma) \):
\[
D_{m', m}^{(j)}(\alpha, \beta, \gamma) = e^{-i m' \alpha} d_{m', m}^{(j)}(\beta) e^{-i m \gamma},
\]
where:
\begin{itemize}
    \item \( j \) is the degree of the spherical harmonic,
    \item \( m \) and \( m' \) are the magnetic quantum numbers, which range from \( -j \) to \( j \),
    \item \( d_{m', m}^{(j)}(\beta) \) is the small Wigner \( d \)-matrix, defined as:
    \begin{align}
    \nonumber d_{m', m}^{(j)}(\beta) = \sum_{k=0}^{j+m} 
    \binom{j+m}{k} \binom{j-m}{j+m-k} \\
    \cos^{2j-k}\left( \frac{\beta}{2} \right) 
    \sin^k\left( \frac{\beta}{2} \right).
    \end{align}
\end{itemize}

\paragraph{SH Coefficient Transformation}
To rotate view-dependent spherical harmonic (SH) coefficients \( \mathcal{C}' \) before simulation, we compute the Wigner D-matrix for a given rotation matrix, derived from Euler angles \( (\alpha, \beta, \gamma) \). Let the view-dependent SH coefficients be denoted as:
\[
\mathcal{C}' = \{ c_m^{(j)} \mid m = -j, -j+1, \dots, j \},
\]
where \( c_m^{(j)} \) corresponds to the coefficient of degree \( j \) and magnetic quantum number \( m \).

For rotation, the transformed SH coefficients \( \hat{\mathcal{C}} \) are computed as:
\[
\hat{\mathcal{C}} = D^{(j)} \mathcal{C}',
\]
where \( D^{(j)} \) is the Wigner D-matrix of degree \( j \). Specifically, each SH coefficient \( c_m^{(j)} \) is rotated using its corresponding Wigner D-matrix element:
\[
\hat{c}_m^{(j)} = \sum_{m'=-j}^{j} D_{m', m}^{(j)}(\alpha, \beta, \gamma) c_{m'}^{(j)}.
\]
Here, \( \hat{c}_m^{(j)} \) is the transformed coefficient, and the sum runs over all values of \( m' \) from \( -j \) to \( j \).

\subsection{Joint Poisson Fields}
To resolve conflicts (Sec. 4.2.1) between the indicator functions \(\mathcal{X}_\mathcal{S}\) and \(\mathcal{X}_\mathcal{O}^\mathcal{S}\), we first ensure that \(\mathcal{S}\) remains smooth and continuous before addressing the conflicts. The process is performed iteratively as follows:

\begin{enumerate}
    \item Identify Surface Points of \(\mathcal{S}\):
    \begin{itemize}
        \item Intersection Region:
        \[
        \{ x \mid 0.5 < \mathcal{X}_\mathcal{S}(x) < 0.6 \text{ and } \mathcal{X}_{\mathcal{O}}^{\mathcal{S}}(x) > 0.5 \}
        \]
        \item Non-Intersection Region:
        \[
        \{ x \mid 0.5 < \mathcal{X}_\mathcal{S}(x) < 0.6 \text{ and } \mathcal{X}_{\mathcal{O}}^{\mathcal{S}}(x) < 0.5 \}
        \]
    \end{itemize}
    
    \item Compute Mean Curvature:
    \begin{itemize}
        \item Compute the mean curvature \(H(x)\) at each surface point using neighboring points.
    \end{itemize}

    \item Adjust Surface Points in Intersection Regions:
    \begin{itemize}
        \item For each point \(x\) in the intersection region, find the nearest point \(x_{\text{closest}}\) in the non-intersection region. If \(|H(x) - H(x_{\text{closest}})| > \tau\), update:
        \[
        \mathcal{X}_\mathcal{S}(x) = 0.49.
        \]
    \end{itemize}

    \item Ensure Surface Smoothness:
    \begin{itemize}
        \item Repeat steps 1–3 for 10 iterations to achieve smoothness.
    \end{itemize}

    \item Resolve Conflicts in \(\mathcal{X}_\mathcal{O}^\mathcal{S}\):
    \begin{itemize}
        \item For points \(x\) where \(\mathcal{X}_\mathcal{O}^\mathcal{S}(x) > 0.5\), check neighbors. If any neighbor satisfies \(\mathcal{X}_\mathcal{S}(x) > 0.5\), update:
        \[
        \mathcal{X}_\mathcal{O}^\mathcal{S}(x) = 0.49.
        \]
    \end{itemize}
\end{enumerate}

\subsection{Normals Disambiguities}
The normals of \(\{\boldsymbol{k_g}\}_{g \in \mathcal{S}}\) (Sec. 4.2.1) correspond to the direction of the minimum scale factor of the flattened Gaussian. Due to ambiguity in determining the normal direction, as both directions along the shortest axis are possible, we resolve this by utilizing the training viewing direction. Specifically, we ensure the angle between the normal and viewing directions exceeds 90 degrees, as observations are made from the exterior of the surface. We then count the occurrences of each direction across training views and select the one with the highest number of votes.
\subsection{Mesh2Gaussians}
We bind new Gaussians to the mesh triangles (Sec. 4.2.4) as follows: for a given triangle with vertices \(\{\boldsymbol v_1, \boldsymbol v_2, \boldsymbol v_3\}\), the center of the new Gaussian is set at the centroid of the triangle, calculated as:
\[
\boldsymbol k = \frac{1}{3}(\boldsymbol v_1 + \boldsymbol v_2 + \boldsymbol v_3).
\]
The normal vector \(\boldsymbol r_1\) to the plane of the triangle is computed as:
\[
\boldsymbol r_1 = \frac{(\boldsymbol v_2 - \boldsymbol k) \times (\boldsymbol v_3 - \boldsymbol k)}{\|(\boldsymbol v_2 - \boldsymbol k) \times (\boldsymbol v_3 - \boldsymbol k)\|},
\]
where \(\times\) denotes the cross product. The second Gaussian axis \(\boldsymbol r_2\) is defined as:
\[
\boldsymbol r_2 = \frac{\boldsymbol v_2 - \boldsymbol k}{\|\boldsymbol v_2 - \boldsymbol k\|}.
\]
The third vector \(\boldsymbol r_3\) is computed through a one-step Gram-Schmidt projection \cite{bjorck1994numerics}:
\[
\boldsymbol r_3 = \text{proj}(\boldsymbol v_3 - \boldsymbol k; \boldsymbol r_1, \boldsymbol r_2).
\]
The Gaussian rotation matrix is then defined as:
\[
\boldsymbol R = [\boldsymbol r_1, \boldsymbol r_2, \boldsymbol r_3].
\]
The scaling values are calculated as follows: \(s_2 = \|\boldsymbol v_2 - \boldsymbol k\|\) for the direction \(\boldsymbol r_2\), \(s_3 = \| \boldsymbol r_3^T(\boldsymbol v_3 - \boldsymbol k )\|\) for the direction \(\boldsymbol v_3\), and \(s_1 = \epsilon\), where \(\epsilon=1\times 10^{-8}\), for the shortest axis \(\boldsymbol r_1\) to account for the flattened 3D Gaussian.

\section{Experiments}
\begin{figure}[!htb]
    \centering
    \includegraphics[width=\linewidth]{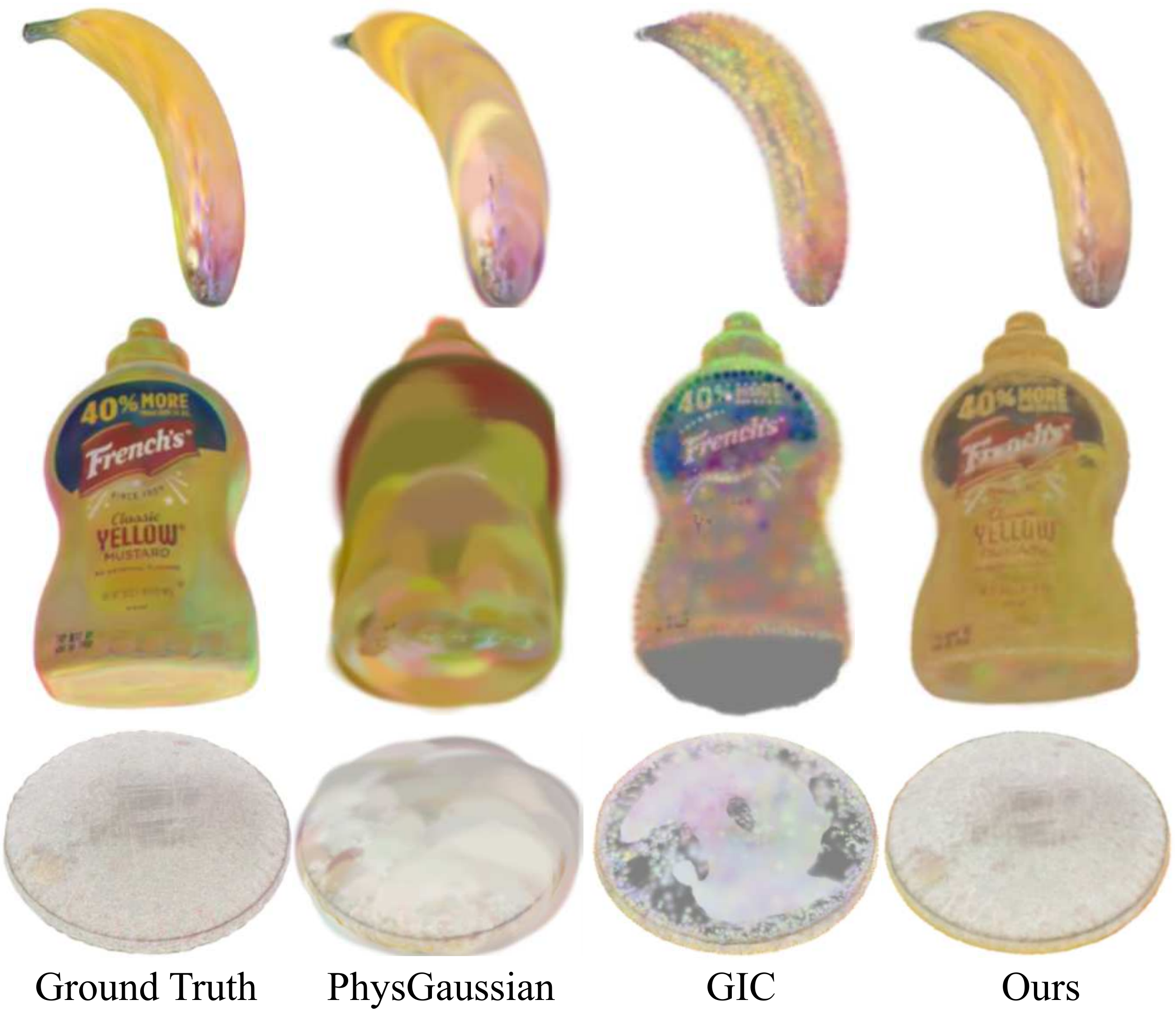}
\caption{\textbf{Object Restoration.} Restored objects from the Decoupling benchmark are rendered from different viewpoints. From top to bottom: Banana, Pillow, and Mustard.}
    \label{fig: object_qualitative}
\end{figure}
\subsection{User Study Statistics}
The statistics for each participant (Sec. 5.2) in our user study on in-the-wild video evaluation are summarized in \cref{tab: comparison}. Notably, all participants rated our method the highest across three tasks: scene restoration, object restoration, and object-scene interactive simulation.
\begin{table*}[!htb]
\centering
\caption{\textbf{User Study Statistics.} U1, U2, ..., U10 represent the IDs of individual participants.}\label{tab: comparison}
\resizebox{\linewidth}{!}{
\begin{tabular}{c|c|c|c|c|c|c|c|c|c|c|c}
\toprule
\multicolumn{1}{c}{Metrics} &\multicolumn{1}{c}{Methods}&\multicolumn{1}{c}{U1}&\multicolumn{1}{c}{U2}&\multicolumn{1}{c}{U3}&\multicolumn{1}{c}{U4}&\multicolumn{1}{c}{U5}&\multicolumn{1}{c}{U6}&\multicolumn{1}{c}{U7}&\multicolumn{1}{c}{U8}&\multicolumn{1}{c}{U9}&\multicolumn{1}{c}{U10}\\
\midrule
\multirow{3}{*}{\makecell{Scene Restoration \\ (SRQ $\uparrow$)}} 
& GScream &1.80&2.00&2.00&2.40&2.20&1.80&2.20&1.60&2.20&1.20\\
& VR-GS &2.00&2.20&2.20&2.40&2.20&2.00&2.60&1.80&2.20&1.60\\
& \textbf{Ours} &\textbf{3.20}&\textbf{3.60}&\textbf{3.60}&\textbf{4.00}&\textbf{3.60}&\textbf{3.20}&\textbf{3.40}&\textbf{3.80}&\textbf{3.60}&\textbf{2.80}\\
\midrule

\multirow{3}{*}{\makecell{Object Restoration \\ (ORQ $\uparrow$)}} 

& PhysGaussian &1.50 &1.50 &1.00& 1.50&1.00&1.25&1.50&1.50&1.50&1.75\\
& GIC &2.00&1.75 &1.50&1.75&1.25&2.00&1.25&2.00&1.50&1.00\\
& \textbf{Ours}&\textbf{4.00}& \textbf{4.50}&\textbf{4.25}&\textbf{4.00}&\textbf{3.25}&\textbf{4.00}&\textbf{3.75}&\textbf{4.25}&\textbf{4.00}&\textbf{4.25}\\
\midrule

\multirow{4}{*}{\makecell{Interactive Simulation  \\ (ISF $\uparrow$)}} 
& VR-GS($\mathcal{S}$)+PhysGaussian($\mathcal{O}$) &
1.50&1.50&1.25&1.75&1.50&1.50&1.25&1.75&1.50&1.50\\
& Ours($\mathcal{S}$)+PhysGaussian($\mathcal{O}$)  &2.50&2.75&2.25&2.50&2.00&3.00&2.50&3.00&2.75&2.75\\
& Ours($\mathcal{S}$)+GIC($\mathcal{O}$) 
&3.00&3.00&2.50&3.00&2.75&2.25&2.75&2.50&2.50&3.00\\
&\textbf{Ours($\mathcal{S}$)+Ours($\mathcal{O}$)} 
&\textbf{4.25}&\textbf{4.50}&\textbf{4.25}&\textbf{4.00}&\textbf{4.50}&\textbf{4.25}&\textbf{4.50}&\textbf{4.25}&\textbf{4.50}&\textbf{4.50}\\
\bottomrule
\end{tabular}}
\end{table*}
\begin{figure}[!htb]
    \centering
    \includegraphics[width=\linewidth]{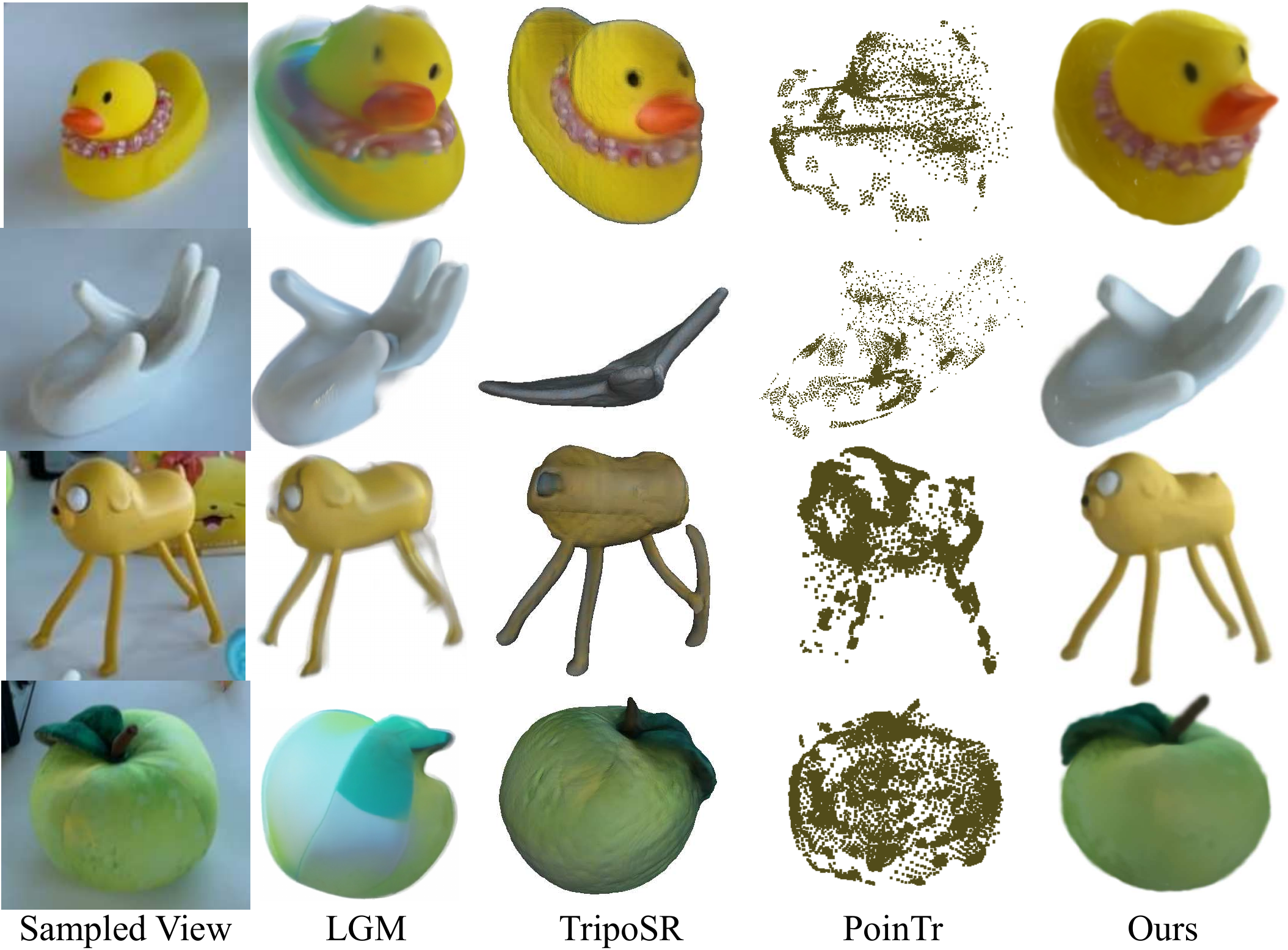}
    \caption{\textbf{Generative Models.} Results for state-of-the-art models using a single image as input: Gaussian generative model LGM \cite{tang2025lgm} and mesh generative model TripoSR \cite{tochilkin2024triposr}; and proxy points as input: point cloud completion model PoinTr \cite{yu2021pointr}.}
    \label{fig: generative_qulitative}
\end{figure}
\begin{figure}[!htb]
    \centering
    \includegraphics[width=\linewidth]{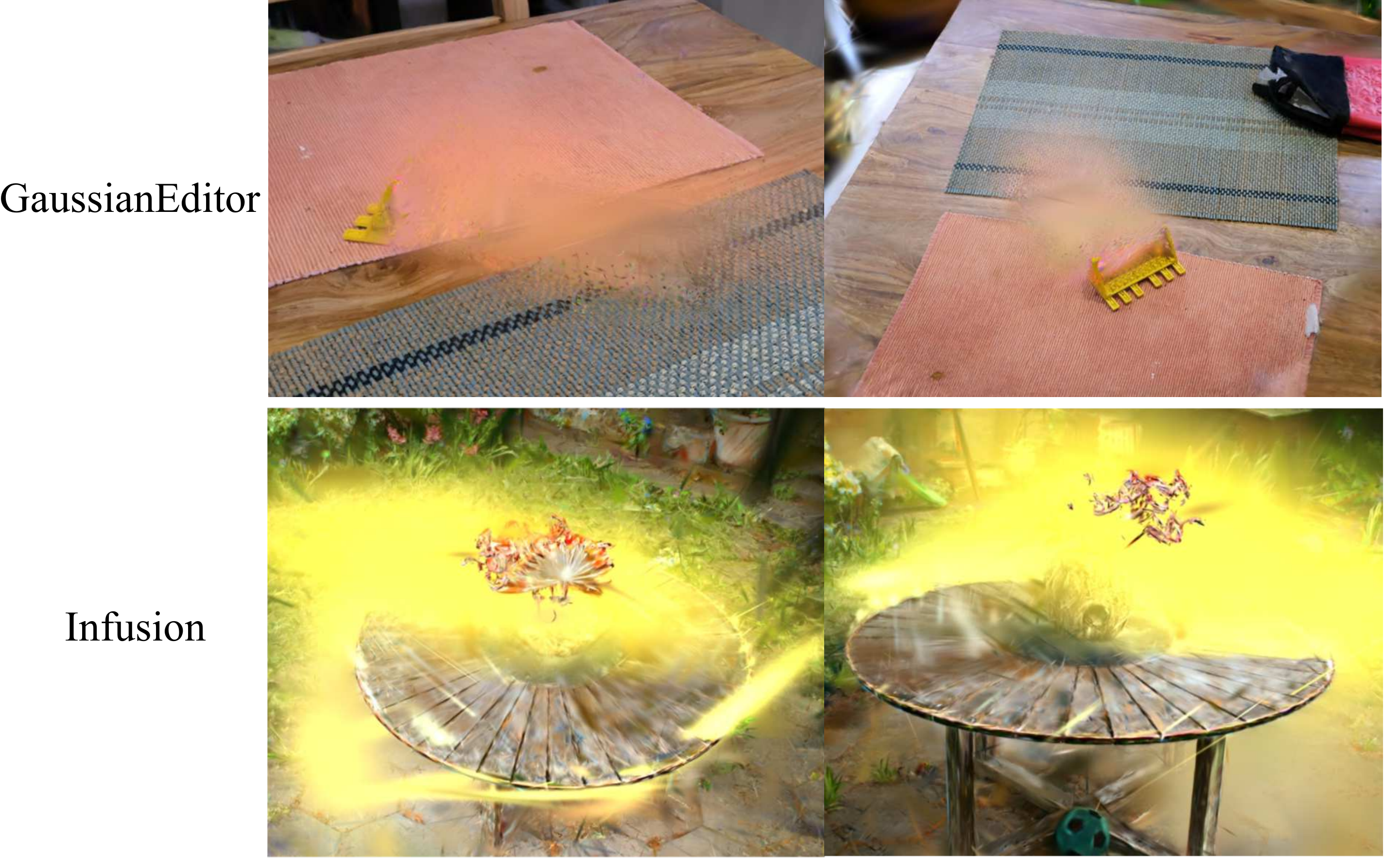}
    \caption{\textbf{Additional Methods.} Results for GaussianEditor \cite{chen2024gaussianeditor} and Infusion \cite{liu2024infusion} reveal significant limitations when using their provided implementations.}
    \label{fig: othermethods}
\end{figure}
\subsection{Additional Evaluations}
\begin{figure*}[!htb]
    \centering
    \includegraphics[width=\linewidth]{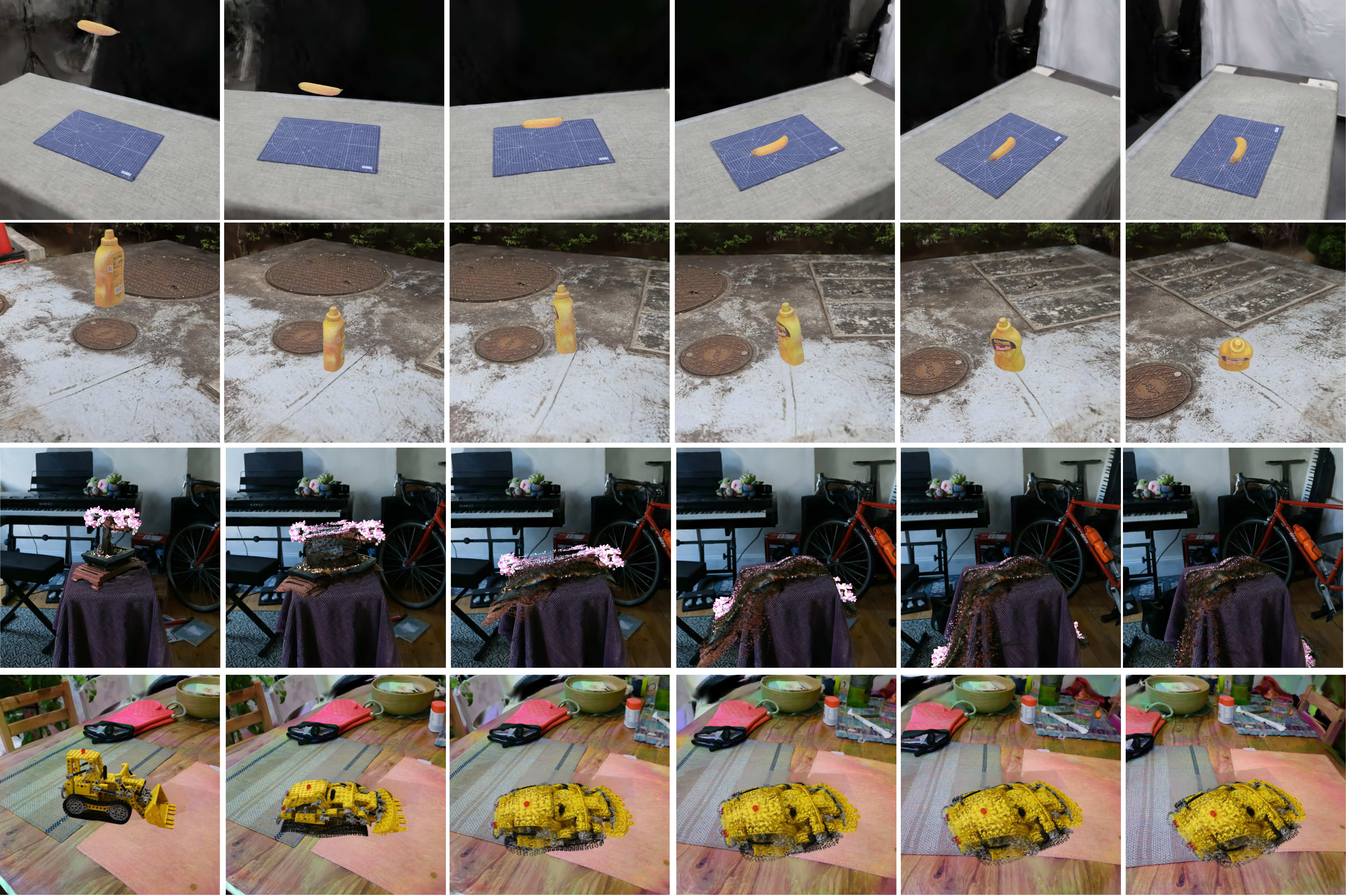}
\caption{\textbf{Additional Interactive Simulations.} Evaluations are rendered with moving cameras. From top to bottom: Banana (collision and elasticity), Mustard (collision and elasticity), Bonsai (fracture and granular flow), and Kitchen (melting and granular flow).}
    \label{fig: simulation_qualitative}
\end{figure*}

\paragraph{Object Restoration} As shown in \cref{fig: object_qualitative}, we present restored objects rendered from various viewpoints derived from the interactive simulation in Fig. 8.  Our joint Poisson field \( \mathcal{W} \) effectively repairs incomplete and broken surfaces of \( \mathcal{O} \), outperforming PhysGaussian \cite{xie2024physgaussian} and GIC \cite{cai2024gaussian}. By bounding the dense points \( P_\mathcal{O} \) within the object's interior, our method achieves superior restoration of both texture and geometry compared to these approaches.
\paragraph{Interactive Simulation} As shown in \cref{fig: simulation_qualitative}, we provide additional qualitative evaluations of our method applied to object-scene interactive simulation, which are not included in the main paper or supplementary video.

\paragraph{Simulator}
Our MLS-MPM implementation leverages both NVIDIA Warp~\cite{warp2022} and Taichi~\cite{10.1145/3355089.3356506}. We performed timing comparisons using the Wolf, Pillow2Sofa, and VaseDeck datasets provided by PhysGaussian~\cite{xie2024physgaussian}. The results are shown in Table~\ref{tab:timing_comparison}, where the computation time per update timestep is expressed in \(10^{-3}\,\text{s}\).

\begin{table}[!htb]
    \centering
    \caption{Timing comparison of MLS-MPM simulation engines. Computation time per update timestep (in \(10^{-3}\,\text{s}\)).}
    \label{tab:timing_comparison}
    \resizebox{\linewidth}{!}{
    \begin{tabular}{lccc}
        \toprule
        \textbf{Method} & \textbf{Wolf} & \textbf{Pillow2Sofa} & \textbf{VaseDeck} \\
        \midrule
        Taichi~\cite{10.1145/3355089.3356506} & 2.560 & \textbf{1.390} & 0.583 \\
        Warp~\cite{warp2022} & \textbf{2.290} & 2.510 & \textbf{0.538} \\
        \bottomrule
    \end{tabular}}
\end{table}
\paragraph{Generative Models} We evaluate state-of-the-art generative models for mesh generation \cite{tochilkin2024triposr,voleti2025sv3d}, Gaussian generation \cite{tang2025lgm}, and point cloud completion \cite{yu2021pointr,kasten2024point}. The first two models take a single frame as input, while the point cloud completion models use our proposed proxy points $\mathcal{P}_\mathcal{O}$ as input. Although some models generate reasonable shapes (see \cref{fig: generative_qulitative}), they often fail for untrained inputs, exhibiting inaccuracies in geometry and texture that diverge from the target properties in the raw scene.

\paragraph{Additional Methods}  
We evaluate two recent approaches, GaussianEditor \cite{chen2024gaussianeditor} and Infusion \cite{liu2024infusion}, for scene \( \mathcal{S} \) restoration. As shown in \cref{fig: othermethods}, both methods exhibit significant limitations. Infusion suffers from severe errors due to inaccurate depth estimation and projection issues in its implementation. Similarly, GaussianEditor demonstrates inconsistent segmentation propagation across views, leading to incomplete object removal or residual artifacts. Additionally, our experiments show that GaussianEditor runs approximately five times slower than the runtime reported in the original paper. These limitations have also been noted by other users on the GitHub issue channels for the respective implementations.

{
    \small
    \bibliographystyle{ieeenat_fullname}
    \bibliography{main}
}